\documentclass[aps,prd,twocolumn,a4paper,superscriptaddress,nofootinbib]{revtex4-1}
\usepackage{graphicx,caption,subcaption,amsmath,amsfonts,amssymb,multirow,extarrows,bm,acronym,float}
\usepackage[colorlinks,linkcolor=blue,citecolor=blue,urlcolor=blue ]{hyperref}
\usepackage[flushleft]{threeparttable}
\usepackage{xcolor, CJK}
\floatstyle{plaintop}
\restylefloat{table}
\newcommand{\Msun}{\,{\rm M}_\odot}

\newcommand{\PKU}{Kavli Institute for Astronomy and Astrophysics, Peking University, Beijing 100871, China}
\newcommand{\NAOCAS}{National Astronomical Observatories, Chinese Academy of Sciences, Beijing 100012, China}

\def\mnras{MNRAS}            
\def\prd{Phys.~Rev.~D}       
\def\prl{Phys.~Rev.~Lett}    

\begin{document}
\begin{CJK*}{UTF8}{gbsn} 

\title{Effect of Noise Estimation in Time-Domain Ringdown Analysis: A Case Study with GW150914}

\author{Hai-Tian Wang (王海天)}
\email[Corresponding author: ]{wanght@pku.edu.cn}
\affiliation{\PKU}
\author{Lijing Shao (邵立晶)}
\email[Corresponding author: ]{lshao@pku.edu.cn}
\affiliation{\PKU}
\affiliation{\NAOCAS}

\date{\today}

\begin{abstract}
Accurate noise estimation from gravitational wave (GW) data is critical for Bayesian inference. 
However, recent studies on ringdown signal, such as those by \citet{2021PhRvL.127a1103I}, \citet{2022PhRvL.129k1102C}, and \citet{2022arXiv220202941I}, have encountered disagreement in noise estimation, leading to inconsistent results. 
The key discrepancy between these studies lies in the usage of different noise estimation methods, augmented by the usage of different sampling rates.
We achieved consistent results across various sampling rates by correctly managing noise estimation, shown in the case study of the GW150914 ringdown signal. 
By conducting a time-domain Bayesian inference analysis on GW data, starting from the peak of the signal, we discovered that the first overtone mode is weakly supported by the amplitude distribution, with a confidence level of $1.6\sigma$, and is slightly disfavored by the log-Bayes factor. Overall, in our time-domain analysis we conclude there is no strong evidence for overtones in GW150914.

\end{abstract}

\maketitle
\end{CJK*}

\acrodef{GW}{gravitational wave}
\acrodef{GWTC-3}{the third Gravitational-wave Transient Catalog}
\acrodef{FFT}{fast Fourier transform}
\acrodef{LIGO}{Laser Interferometer Gravitational-Wave Observatory}
\acrodef{LVC}{LIGO-Virgo Collaboration}
\acrodef{LVK}{LIGO-Virgo-KAGRA}
\acrodef{BNS}{binary neutron star}
\acrodef{NR}{numerical relativity}
\acrodef{BH}{black hole}
\acrodef{BBH}{binary black hole}
\acrodef{GR}{general relativity}
\acrodef{SNR}{signal-to-noise ratio}
\acrodef{PSD}{power spectral density}
\acrodef{PDF}{probability density function}
\acrodef{ACF}{autocovariance function}
\acrodef{IMR}{inspiral-merger-ringdown}
\acrodef{QNMs}{quasinormal modes}

\section{Introduction}\label{sec:intro}

Over $90$ \ac{GW} events have been identified by the \ac{LVK} Collaboration throughout its first three observing runs \citep{LIGO_PRX2019,LIGO_O3a_PRX2020,2021arXiv211103606T}.
It is envisaged that the aftermath of a violent collision is a distorted object that emits ringdown signal during its oscillatory phase until it stabilizes \citep{Hawking:1971vc,PhysRevLett.34.905}. 
The ringdown signal is characterized by a superposition of \ac{QNMs} \citep{Schw_PRD_Vishveshwara1970,GW_APJL_Press1971,QNM_APJ_Teukolsky1973}, which are usually decomposed into spin-weighted spheroidal harmonics with angular indices $(\ell,m)$. 
Each angular index encompasses a series of overtone modes, represented by $n$ \citep{Berti:2009kk}.
The analysis of ringdown signal presents a unique opportunity to test \ac{GR} in the strong field of gravity.

Typically, aside from the fundamental mode $(\ell=m=2,n=0)$, we anticipate that higher multipoles would be readily detectable in GW data for asymmetric mass-ratio systems \citep{Berti:2007zu,Gossan:2011ha,London:2014cma,Brito:2018rfr}.
However, for events similar to GW150914 \citep{gw150914_PRL2016}, which has a mass ratio nearing $1$, the contributions of higher multipoles can be disregarded.
Note that higher modes excitation is also strongly correlated with the source inclination.
Contributions of them are suppressed for a face-off source, which is the case for GW150914 \citep{LIGOScientific:2016vpg}.
GW150914 represents the first \ac{BBH} event identified by the \ac{LVK} Collaboration during its first two observing runs \citep{LIGO_PRX2019}.
This event is particularly suitable for ringdown analyses, given that the \ac{SNR} contained in the ``linear" regime is around $8$ while the post-peak SNR is around $14$ \citep{2019PhRvL.123k1102I,2021PhRvD.103l2002A,2021arXiv211206861T,Test_GR_150914}. Note that the post-peak data are likely not entirely describable through a linear ringdown model, because nonlinear and time-dependent corrections have a big impact very close to the peak~\cite{Baibhav:2023clw}
\citet{PRD_Carullo2019} performed an analysis of its ringdown signal and confirmed the absence of evidence for higher multipoles.

Promisingly, \citet{Overtone_PRX_Giesler2019} determined that when overtone modes are incorporated into the ringdown waveform, it can correspond to a \ac{NR} waveform commencing from the peak amplitude, immediately following the merger. 
The existence of the first overtone mode, based on GW150914, has been investigated by various methods.
These include the time-domain (TD) method \citep{2021PhRvL.127a1103I,2022PhRvL.129k1102C,2021PhRvD.103l2002A,2021arXiv211206861T,Crisostomi:2023tle,Isi:2023nif,Carullo:2023gtf}, the frequency-domain (FD) method \citep{2021PhRvD.104l3034F,2022PhRvD.106d3005F,2021PhRvD.103b4041B,Wang:2023xsy}, and the mode cleaning method \citep{2022PhRvD.106h4036M,Ma:2023vvr,Ma:2023cwe}.
Among these techniques, the TD method is particularly prevalent and has been applied in ringdown analyses for other events \citep{2021PhRvD.103l2002A,2021arXiv211206861T}.
Moreover, it has been extensively utilized in testing the no-hair theorem \citep{2019PhRvL.123k1102I,2021PhRvD.103b4041B}, the \ac{BH} area law \citep{2021PhRvL.127a1103I}, non-Kerr parameters \citep{2021PhRvD.103l2002A,2021arXiv211206861T,2021PhRvD.104j4063W,Cheung:2020dxo,Mishra:2021waw,Carullo:2021dui,Dey:2022pmv,Carullo:2021oxn,Gu:2023eaa,Laghi:2020rgl}, as well as in the exploration of \ac{BH}  thermodynamics \citep{Hu:2021lbt,Carullo:2021yxh}.

However, employing the TD method, \citet{2022PhRvL.129k1102C} concluded that the ``{\it claims of an overtone detection are noise dominated}'' when they scrutinized the ringdown signal of GW150914 at a sampling rate of $16$ kHz.
This is at odds with the findings of \citet{2021PhRvL.127a1103I}, who analyzed the identical ringdown signal at a sampling rate of $2$ kHz.
In response to \citet{2022PhRvL.129k1102C}, \citet{2022arXiv220202941I} reanalyzed the same ringdown signal using different solutions for the TD method (i.e., a distinct sampling algorithm, sampling rate at $4$ kHz, and a Fourier based autocorrelation function estimation method), leading to different parameter constraints compared to \citet{2022PhRvL.129k1102C} and \citet{2021PhRvL.127a1103I}.
A crucial aspect of these solutions is the autocorrelation function (ACF) estimation method.
As explained by \citet{2021arXiv210705609I}, it can be computed directly from the GW data, or it can be truncated from the inverse \ac{FFT} of the one-sided \ac{PSD} in accordance with the Wiener-Khinchin theorem.
We referred to these two TD methods as the TTD1 method and the TTD2 method, respectively.
The default method in pyRing, hence in e.g. Refs.~\citep{2021PhRvD.103l2002A,2021arXiv211206861T} is TTD2.
The TTD1 method was used in Refs.~\citep{2021PhRvL.127a1103I, 2022PhRvL.129k1102C}.
The TTD2 method is expected to be more robust, and we will employ it to analyze the ringdown signal of GW150914.

The organization of this paper is as follows. In Sec.~\ref{sec:noise}, we present comparisons between noise estimates. 
In Sec.~\ref{sec:bayes}, we show the results of Bayesian inference using the TD method. 
Finally, in Sec.~\ref{sec:conclusion}, we provide a succinct summary and discussion. 
Unless specified otherwise, we adopt geometric units with $G=c=1$ throughout the paper.

\section{The comparison of noise estimates}\label{sec:noise}

Within the framework of \ac{GR}, the TD ringdown waveform of a Kerr \ac{BH} can be represented as
\begin{equation}
\begin{aligned}
&h_{+}(t)-ih_{\times}(t) \\
=&\sum_{\ell}^{}\sum_m^{}\sum_n^{N}A_{\ell mn}\exp(i2\pi f_{\ell mn}t+i\phi_{\ell mn}-\frac{t}{\tau_{\ell mn}}) \\
& \times{}_{-2}Y_{\ell m}(\iota, \delta).
\label{eq:rin_td}
\end{aligned}
\end{equation}
Here, $N$ denotes the total number of the overtone modes under consideration, while $A_{\ell mn}$ and $\phi_{\ell mn}$ correspond to the amplitudes and phases for the various modes.
$\iota$ and $\delta$ represent the inclination and azimuth angles, respectively, with the latter fixed at zero for our study.
It should be noted that we disregard the contributions from higher multipoles due to the absence of evidence in GW150914 \citep{PRD_Carullo2019}.
$f_{\ell mn}$ denotes the oscillation frequency, and $\tau_{\ell mn}$ the damping time, with both being determined by the final mass ($M_f$) and the final spin ($\chi_f$) of the remnant.
For GW150914, we only consider the $\ell=|m|=2$ multipole and assume $h_{\ell m}=(-1)^{\ell}h^*_{\ell-m}$.
Contributions from mode-mixing are not considered in our analysis.

The ringdown signal is veiled within the noise present in GW data.
To extract the information from the ringdown signal, an understanding of the noise is required.
For these data, two sampling rates are available, $4096$ Hz and $16384$ Hz, provided by the GW Open Science Center (GWOSC).
Before we proceed to estimate the \ac{PSD}, two critical steps must be undertaken.
Firstly, we must resample the GW data to the required sampling rate, which in our case is $2048$ Hz.
Secondly, a high-pass filter can be implemented at approximately $20$ Hz on the resampled data.
If these steps are not appropriately managed, one may end up with biased \ac{PSD}s.
To address the first step, we employ a resampling algorithm that uses the Butterworth filter.
For the second step, we utilize a Finite Impulse Response filter \citep{KhanFIR2020}, setting the order at $512$.
Both of these steps are implemented using {\sc PyCBC} \citep{2012PhRvD..85l2006A}.

\begin{figure}
\centering
\includegraphics[width=0.46\textwidth,height=6cm]{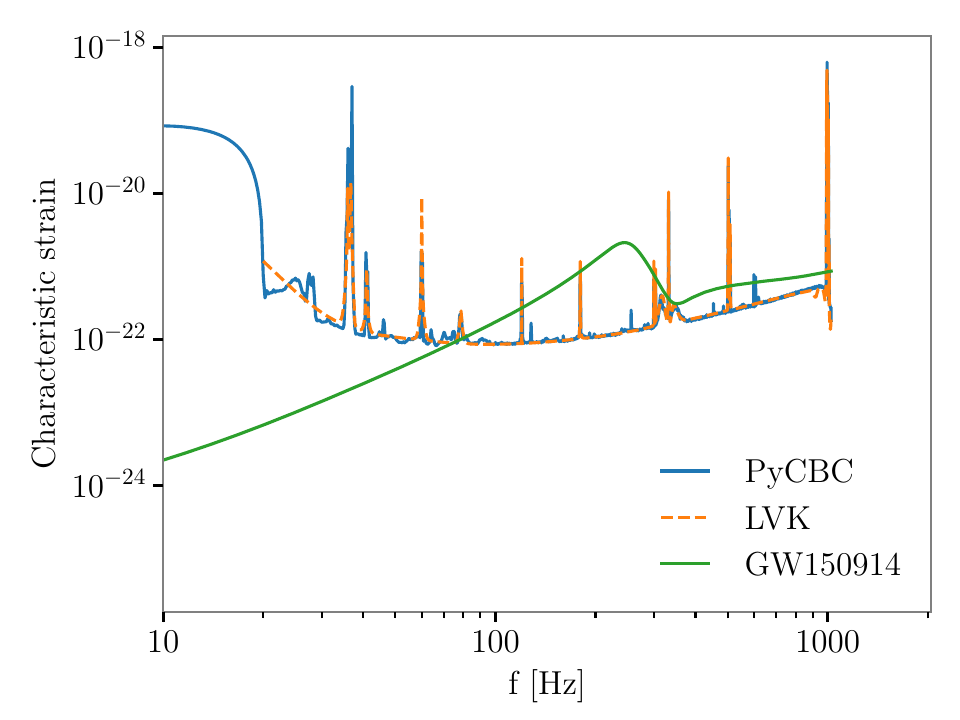}
\caption{
Characteristic strains of a ringdown signal analogous to GW150914 and \ac{PSD}s estimated by various methods.
The dashed green curve is publicly available in Ref.~\cite{psd:gwtc1}.
}\label{fig:psds}
\end{figure}

We assume the noise data are Gaussian and stationary.
Typically, one can estimate the one-side \ac{PSD} using the Welch method \citep{1967D.Welch}. 
In our study, we merge the Welch method with the {\it inverse spectrum truncation} algorithm, as implemented in {\sc PyCBC}.
Moreover, an algorithm based on Bayesian inference, called {\sc BayesLine}, has been developed to model the \ac{PSD} \citep{Littenberg:2014oda,Cornish:2014kda}.
In Fig.~\ref{fig:psds}, we exhibit \ac{PSD}s of GW data detected by the Handford detector, estimated by these various methods.
For the {\sc PyCBC} estimation, we utilize the GW data obtained from the GWOSC \citep{data:gw150914} with a sampling rate of $4096$ Hz and a duration of $4096$ s.
As shown, the \ac{PSD} estimated by the {\sc PyCBC} package aligns closely with that provided by \ac{LVK} Collaboration \citep{psd:gwtc1}.

To gain a deeper understanding of the effects of noise estimation, we also give a characteristic strain of a ringdown signal similar to GW150914.
The definition of characteristic strain can be found in Ref.~\citep{Moore:2014lga}.
We inject a GW150914-like ringdown signal into the Hanford detector, characterized by the following parameters:
$M_f=71.73 \, \Msun$, $\chi_f=0.74$, $A_{220}=0.73\times 10^{-20}$, $A_{221}=0.95\times 10^{-20}$, $\phi_{220}=0.95$, $\phi_{221}=2.28$, $\iota=2.42$, ${\rm RA}=1.95$, ${\rm DEC}=-1.27$, $\psi=0.82$, where ${\rm RA}$ and ${\rm DEC}$ represent two sky position angles and $\psi$ denotes the polarization angle.
The detected ringdown signal can be expressed as $h(t)=F^+h_++F^{\times}h_{\times}$, where $F^{+,\times}$ are the antenna pattern functions, determined by the sky location and the polarization angle.
In Fig.~\ref{fig:psds}, we display the dimensionless strain amplitude, $2f|\tilde{h}(f)|$, where $\tilde{h}(f)$ denotes the \ac{FFT} of $h(t)$.

\section{Results of Bayesian inferences}\label{sec:bayes}

To obtain ringdown parameters from GW data $h$, we use an algorithm grounded in the Bayes theorem, $P(\theta|d,I)=P(d|\theta,I)P(\theta|I)/P(d|I)$, where $P(\theta|d,I)$ is the desired posteriors, $P(d|\theta,I)$ is the likelihood function, $P(\theta|I)$ is the priors, $P(d|I)$ is the evidence, $I$ is the chosen model, and $\theta$ stands for the model parameters.
In TD, the log-likelihood function can be expressed as
\begin{equation}
\log \mathcal{L}=-\frac{1}{2}(d(t)-h(t))\mathcal{C}^{-1}(d(t)-h(t))^{\intercal}+C_0, 
\end{equation}
where $\mathcal{C}$ is the autocovariance matrix and $C_0$ is a constant.
In our analysis, the autocovariance matrix adopts the Toeplitz form of the truncated \ac{ACF}.

For the priors of the ringdown parameters, we fix the sky location, polarization angle, and geocentric time, with ${\rm RA}=1.95$, ${\rm DEC=-1.27}$, $\psi=0.82$, and $t_c=t_{\rm ref}+\Delta t$\,s;
$t_{\rm ref}=1126259462.40854$ corresponds to a trigger time at $t_{\rm H1}=1126259462.42323$\,s for the Hanford detector.
For all other parameters, we assume flat priors within the following ranges: $M_f\in[50,100]\Msun$, $\chi_f\in[0,0.99]$, $\cos\iota\in[-1,1]$, $A_{22n}\in[0,5]\times 10^{-20}$, and $\phi_{22n}\in[0,2\pi]$.

\begin{figure}
\centering
\includegraphics[width=0.48\textwidth,height=9cm]{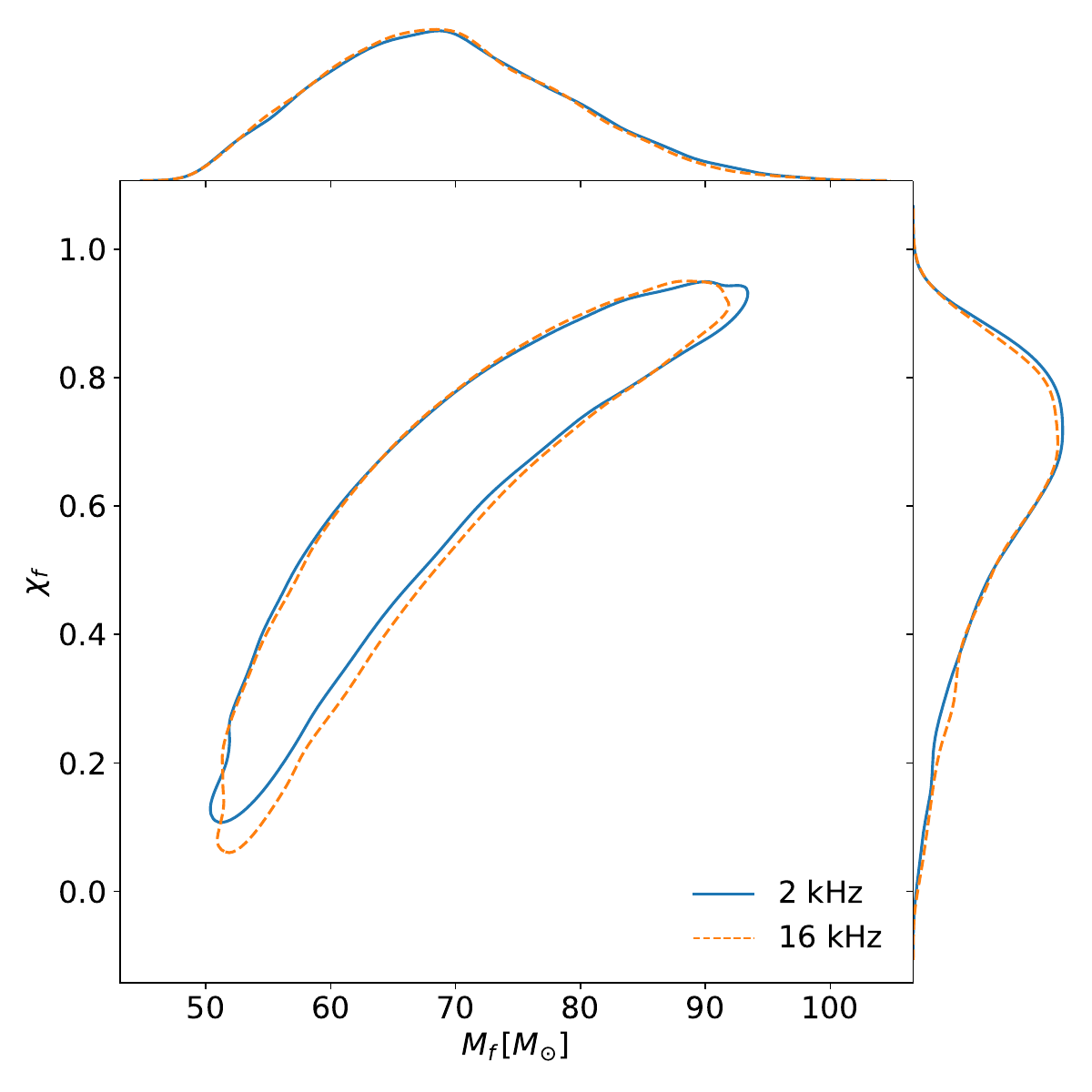}
\caption{
The posterior distributions of the redshifted final mass $M_f$ and final spin $\chi_f$ of the GW150914 remnant, derived using \ac{PSD}s estimated by the {\sc PyCBC} package.
The ringdown signal of GW150914 is analyzed at two distinct sampling rates, specifically, $2048$ Hz and $16384$ Hz.
The contribution of the first overtone mode in the ringdown signal is considered in all instances.
For both analyses, the data commence at the peak strain ($\Delta t=0$).
The contours depict the $90\%$-credible regions for the remnant parameters, while the top and right-hand panels show the one-dimensional ($1$D) posteriors for $M_f$ and $\chi_f$, respectively.
}\label{fig:mfsf1}
\end{figure}

To examine the impact on the parameter estimation of the GW150914 ringdown signal, we perform Bayesian inferences using \ac{ACF}s computed via {\sc PyCBC} with two distinct sampling frequencies, $2048$ Hz and $16384$ Hz.
In the case of a sampling rate of $2048$ Hz, we downsample the GW data from the original raw data, which has a duration of $4096$\,s and a sampling rate of $4096$ Hz.
No downsampling is required for the $16384$ Hz case.
The \ac{PSD}s for both cases are derived from the entire on-source data spanning $4096$\,s.
For the $2048$ Hz ($16384$ Hz) case, the slice duration for the Welch method is $8$ ($2$)\,s, while the data duration used in the likelihood computation is $0.4$ ($0.1$)\,s.
We use different durations for various sampling rates due to two primary reasons. 
Firstly, we aim to maintain a similar dimension for the covariance matrix across varying sampling rates.
Secondly, a duration of $0.5$ ($0.1$)\,s was employed in Refs.\citep{2021PhRvL.127a1103I,2022PhRvL.129k1102C}. We aimed to compare our results with theirs and hence adopted similar settings.
We integrate our algorithm with the {\sc Bilby} package \citep{Ashton_APJ2019} and perform Bayesian inferences utilizing the {\sc dynesty} sampler \citep{Dynesty_MNRAS_Speagle2020}, deploying $1000$ live points and a maximum of $1000$ Markov chain (MC) steps.

We plot the posterior distributions of the redshifted final mass $M_f$ and final spin $\chi_f$ in Fig.~\ref{fig:mfsf1}.
With different sampling frequencies, results of \ac{PSD}s estimated using the {\sc PyCBC} package show negligible differences.
For the $2$ kHz ($16$ kHz) case, the constraints are $M_f=68.8^{+17.3}_{-14.2} \, \Msun$ ($M_f=68.5^{+16.7}_{-13.9} \, \Msun$) and $\chi_f=0.66^{+0.21}_{-0.41}$ ($\chi_f=0.65^{+0.22}_{-0.43}$) at $90\%$ credible level, respectively.
Repeating the analysis using a rate of $4096$ Hz or $8192$ Hz left our conclusions unaltered, utilizing the same noise estimation method.
From these comparisons, we conclude that it is critical to accurately estimate noise to ensure the stability in TD analyses across different sampling rates.

\begin{figure*}
\centering
\begin{subfigure}[b]{0.48\linewidth}
\centering
\includegraphics[width=\textwidth,height=9cm]{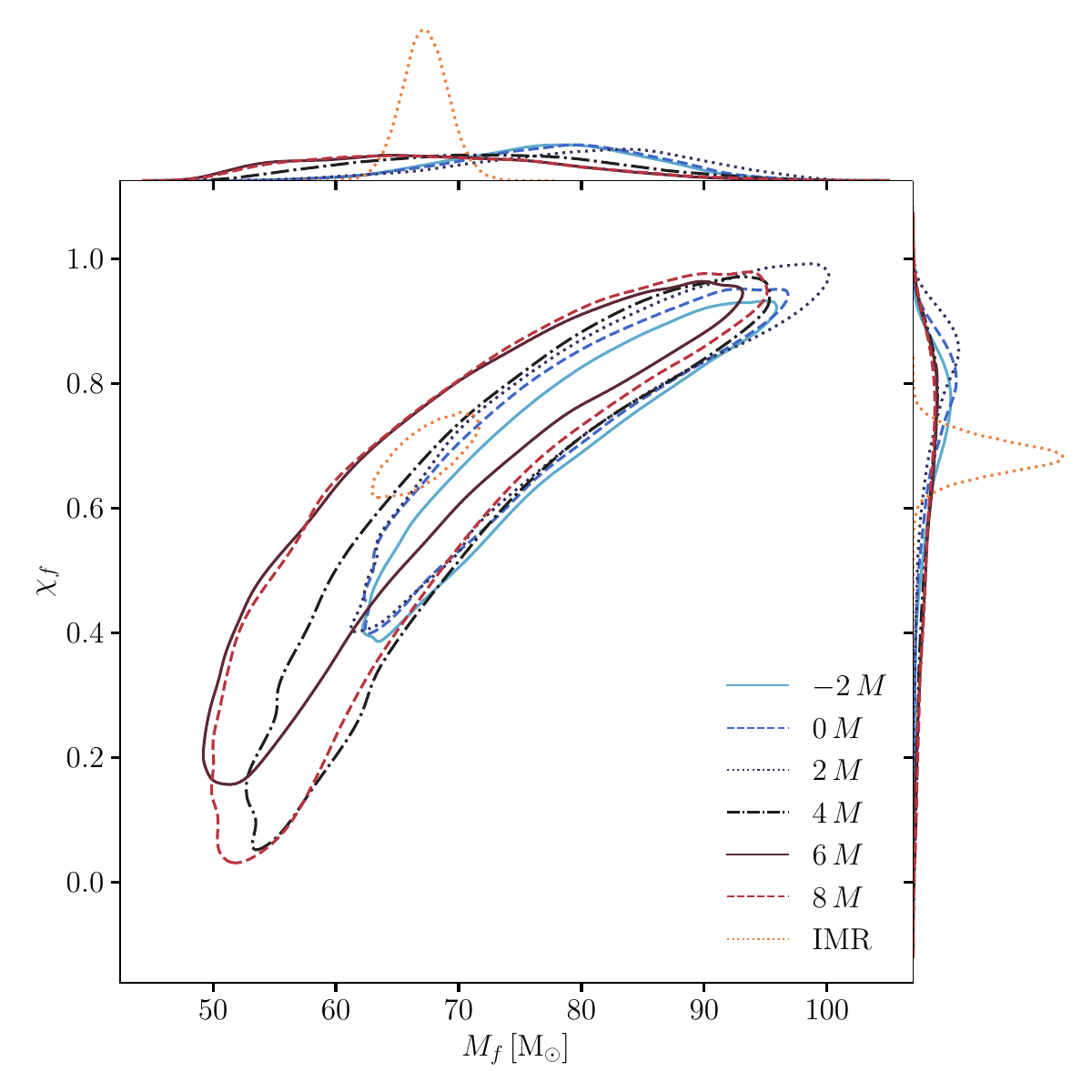}
\end{subfigure}%
\begin{subfigure}[b]{0.48\linewidth}
\centering
\includegraphics[width=\textwidth,height=9cm]{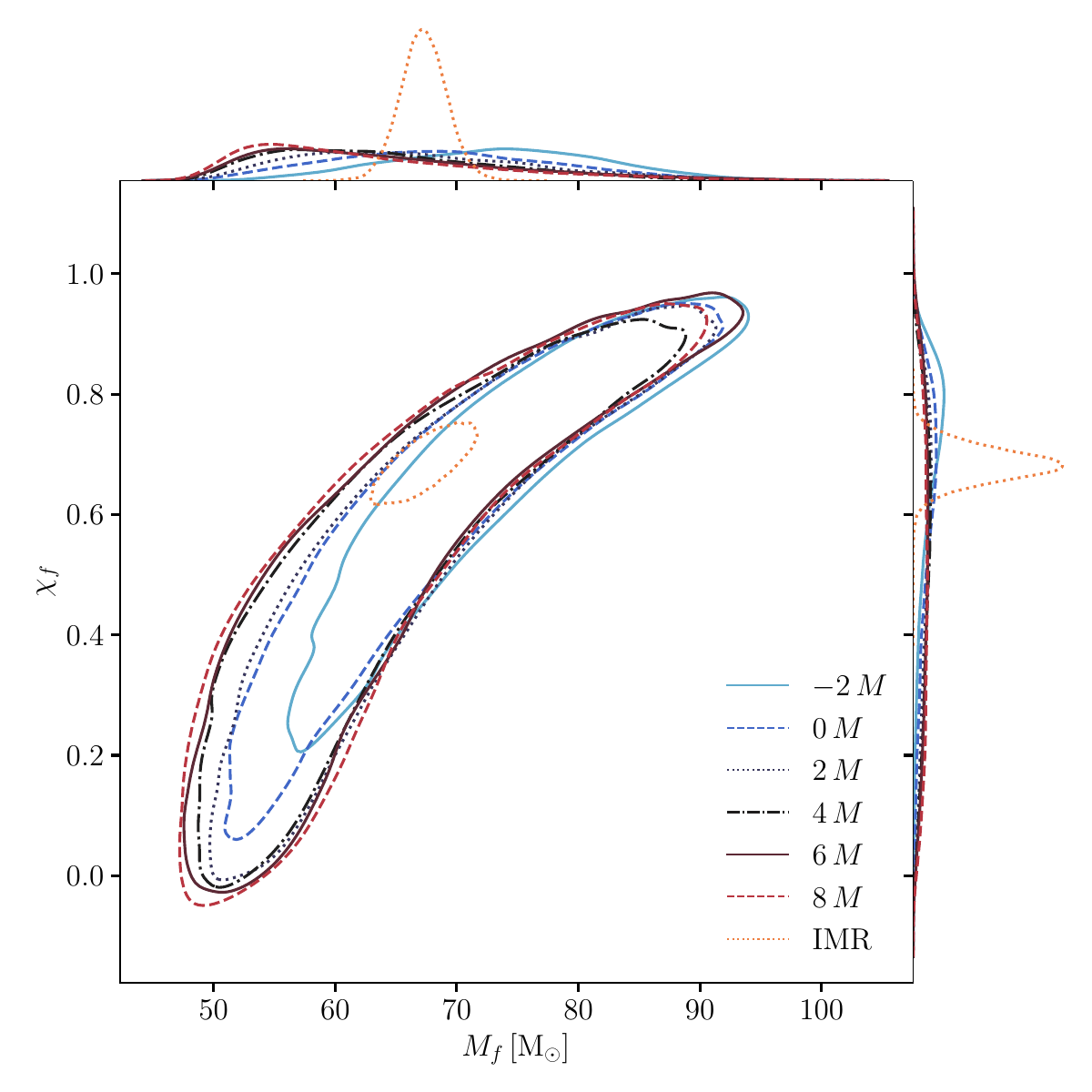}
\end{subfigure}%
\caption{
The posterior distributions of the redshifted final mass ($M_f$) and final spin ($\chi_f$) of the GW150914 remnant, using \ac{PSD}s estimated by the {\sc PyCBC} package.
The contours illustrate $90\%$ credible regions for the remnant parameters, while the top and right-hand panels display $1$D posteriors for $M_f$ and $\chi_f$, respectively.
The left (right) panel denotes results that assume only the fundamental mode (both the fundamental mode and the first overtone mode) in the ringdown signal.
In each panel, the GW data utilized in Bayesian inferences start at different peak times: $\Delta t=(-2,0,2,4,6,8)M$, with $M=68.8\Msun$.
The dotted orange contour indicates the results from the full \ac{IMR} analysis \citep{2021PhRvD.103l2002A}.
}\label{fig:mfsf2}
\end{figure*}

To further scrutinize the presence of the first overtone mode and the stability of the TD method based on the \ac{PSD} estimated by the {\sc PyCBC} package, we conduct an analysis on the ringdown signal using varied starting times, i.e., $\Delta t=(-2,0,2,4,6,8)M$ with $M=68.8\Msun$, and two overtone numbers, $N=(0,1)$.
For those scenarios involving the contribution of the first overtone mode, an inclusion of more GW data (i.e. smaller $\Delta t$ values) in the analyses prompts the posterior distributions of the redshifted final mass to progressively shift towards the high mass region, while the posterior distributions of the final spin gradually transition into the high spin region, as depicted in Fig.~\ref{fig:mfsf2}.
Specifically, when $\Delta t>4M$, the constraints derived from the $N=1$ scenario are weaker than those from the $N=0$ scenario.
Because the \ac{GW} frequency of an \ac{IMR} waveform is a monotonically increasing function of time.
The increase of the frequency with time, which is larger than the difference between the first overtone and the fundamental mode frequency, explains the higher mass obtained. Similar considerations (i.e. a longer amount of signal included) explain the longer $\tau$ hence the larger spin.

\begin{figure}
\centering
\includegraphics[width=0.46\textwidth,height=9cm]{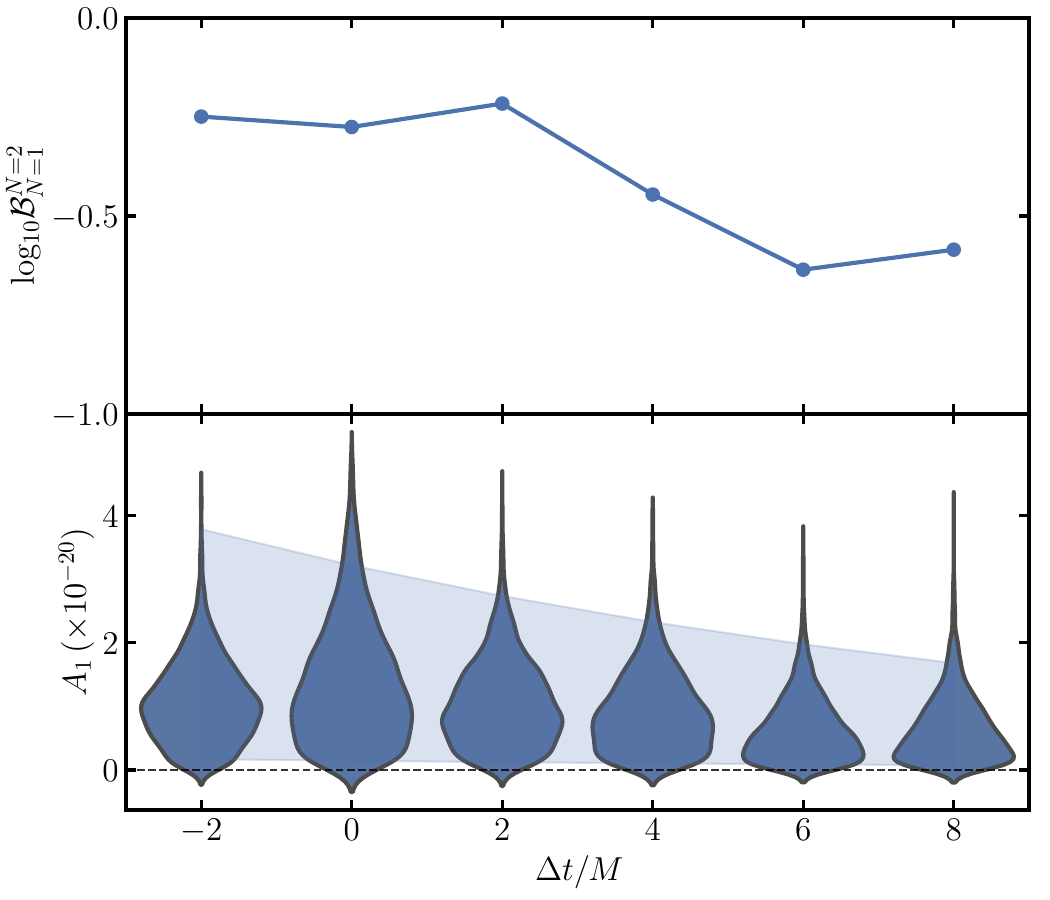}
\caption{
In the upper panel, the log-Bayes factors between the $N=1$ scenario and the $N=0$ scenario are displayed, taking into account various starting times denoted by $\Delta t$.
The lower panel illustrates the distributions of the amplitude of the first overtone mode.
The light blue bands, representing the expected decay rates of $A_1$ starting from $\Delta t=0$, indicate the $90\%$ credible level. \ac{PSD} is estimated by the {\sc PyCBC} package.
}\label{fig:bfs}
\end{figure}

In scenarios where only the fundamental mode is considered, the joint posterior distributions barely cover the median value from the \ac{IMR} analysis until $\Delta t$ exceeds $4M$.
These patterns may be due to the presence of nonlinear signal components close to the merger region \citep{Baibhav:2023clw}.
This observation is consistent with the trend of the log-Bayes factors depicted in Fig.~\ref{fig:bfs}, where a noticeable decline is seen between $\Delta t=2M$ and $\Delta t=6M$.
The fact that log-Bayes factors for all differing $\Delta t$ values are less than zero implies that there is no evidence supporting the first overtone mode.
Conversely, the distributions of the amplitude of the first overtone mode progress as anticipated when $\Delta t>0$.
However, in the case where $\Delta t=0$, the mean of the posteriors of $A_{221}$ is only $1.6\sigma$ from zero.
Therefore, the evidence of the first overtone mode in GW150914 is statistically not significant.
For scenarios where only the contribution of the fundamental mode is considered and it is assumed to start from $\Delta t=8M$, constraints on the remnant parameters are $M_f=67.5^{+18.8}_{-14.1} \, \Msun$ and $\chi_f=0.65^{+0.24}_{-0.46}$ at the $90\%$ credible level.

\section{Discussion and Conclusion}\label{sec:conclusion}

Utilizing the \ac{ACF} based on the {\sc PyCBC} package, we conducted a reanalysis of the GW150914 ringdown signal. 
We highlight that the results remained robust across different sampling rates. 
Then, we perform more analyses, starting from the peak amplitude with a deviation of $\Delta t/M=(-2,0,2,4,6,8)$. 
For the case where $\Delta t=0$, the constraints on the redshifted final mass and the final spin are $M_f=68.5^{+16.7}_{-13.9} \, \Msun$ and $\chi_f=0.65^{+0.22}_{-0.43}$ at the $90\%$ credible level, respectively. 

Additionally, our analysis yielded a log-Bayes factor of $\log\mathcal{B}^{N=1}_{N=0}=-0.2$ when $\Delta t=0$, which remains negative across all $\Delta t$ values. 
However, the log-Bayes factor is affected by the larger parameter space resulting from the additional overtone mode. 
Normally, we also evaluate the evidence of the first overtone mode using its amplitude distribution. 
By this measure, the evidence for the first overtone mode is $1.6\sigma$. 
This is significantly lower than the $3.6\sigma$ reported by \citet{2021PhRvL.127a1103I}.
Thus, the signal strength from the traditional TD method currently employed is not sufficient to confirm a significant contribution from the first overtone mode. 

Multiple studies \citep{Overtone_PRX_Giesler2019,2022PhRvD.106h4036M} have confirmed that the inclusion of higher overtone modes can result in more precise constraints on the parameters of the remnant. 
However, there are some studies demenstrate that it is just a numerical match rather than a physical match.
For example, Refs.~\citep{Baibhav:2023clw,Nee:2023osy,Khera:2023lnc} definitively disproves the physical validity of the ``overtone" model employed in the model under discussion. 

The results of this work might affect data analysis for gravitational waves, including future detectors like the Einstein Telescope \citep{2010CQGra..27s4002P}, Cosmic Explorer \citep{2019BAAS...51g..35R}, Laser Interferometer Space Antenna \citep{LISA_arxiv2017}, TianQin \citep{TQ_2015,TianQin:2020hid}, and Taiji \citep{1093nsrnwx116}.
We use the \texttt{PyCBC} version $2.0.5$ and the \texttt{Bilby} version $1.2.1$.
To allow for reproducibility, we release codes for noise estimation based on the \texttt{PyCBC} package at \citep{acf_estimation_ttd2}.

\begin{acknowledgments}
We thank the anonymous referee for carefully reading the manuscript.
This work was supported by the China Postdoctoral Science Foundation (2022TQ0011), the National Natural Science Foundation of China (12247152, 11975027, 11991053), the National SKA Program of China (2020SKA0120300), the Max Planck Partner Group Program funded by the Max Planck Society, and the High-performance Computing Platform of Peking University.
HTW is supported by the Opening Foundation of TianQin Research Center. 

This research has made use of data or software obtained from the Gravitational Wave Open Science Center (gwosc.org), a service of LIGO Laboratory, the LIGO Scientific Collaboration, the Virgo Collaboration, and KAGRA~\cite{LIGOScientific:2023vdi}. 
LIGO Laboratory and Advanced LIGO are funded by the United States National Science Foundation (NSF) as well as the Science and Technology Facilities Council (STFC) of the United Kingdom, the Max-Planck-Society (MPS), and the State of Niedersachsen/Germany for support of the construction of Advanced LIGO and construction and operation of the GEO600 detector. 
Additional support for Advanced LIGO was provided by the Australian Research Council. 
Virgo is funded, through the European Gravitational Observatory (EGO), by the French Centre National de Recherche Scientifique (CNRS), the Italian Istituto Nazionale di Fisica Nucleare (INFN) and the Dutch Nikhef, with contributions by institutions from Belgium, Germany, Greece, Hungary, Ireland, Japan, Monaco, Poland, Portugal, Spain.
KAGRA is supported by Ministry of Education, Culture, Sports, Science and Technology (MEXT), Japan Society for the Promotion of Science (JSPS) in Japan; National Research Foundation (NRF) and Ministry of Science and ICT (MSIT) in Korea; Academia Sinica (AS) and National Science and Technology Council (NSTC) in Taiwan of China.
\end{acknowledgments}

\bibliographystyle{apsrev4-1}
\bibliography{ttd2}

\begin{thebibliography}{66}%
\makeatletter
\providecommand \@ifxundefined [1]{%
 \@ifx{#1\undefined}
}%
\providecommand \@ifnum [1]{%
 \ifnum #1\expandafter \@firstoftwo
 \else \expandafter \@secondoftwo
 \fi
}%
\providecommand \@ifx [1]{%
 \ifx #1\expandafter \@firstoftwo
 \else \expandafter \@secondoftwo
 \fi
}%
\providecommand \natexlab [1]{#1}%
\providecommand \enquote  [1]{``#1''}%
\providecommand \bibnamefont  [1]{#1}%
\providecommand \bibfnamefont [1]{#1}%
\providecommand \citenamefont [1]{#1}%
\providecommand \href@noop [0]{\@secondoftwo}%
\providecommand \href [0]{\begingroup \@sanitize@url \@href}%
\providecommand \@href[1]{\@@startlink{#1}\@@href}%
\providecommand \@@href[1]{\endgroup#1\@@endlink}%
\providecommand \@sanitize@url [0]{\catcode `\\12\catcode `\$12\catcode
  `\&12\catcode `\#12\catcode `\^12\catcode `\_12\catcode `\%12\relax}%
\providecommand \@@startlink[1]{}%
\providecommand \@@endlink[0]{}%
\providecommand \url  [0]{\begingroup\@sanitize@url \@url }%
\providecommand \@url [1]{\endgroup\@href {#1}{\urlprefix }}%
\providecommand \urlprefix  [0]{URL }%
\providecommand \Eprint [0]{\href }%
\providecommand \doibase [0]{http://dx.doi.org/}%
\providecommand \selectlanguage [0]{\@gobble}%
\providecommand \bibinfo  [0]{\@secondoftwo}%
\providecommand \bibfield  [0]{\@secondoftwo}%
\providecommand \translation [1]{[#1]}%
\providecommand \BibitemOpen [0]{}%
\providecommand \bibitemStop [0]{}%
\providecommand \bibitemNoStop [0]{.\EOS\space}%
\providecommand \EOS [0]{\spacefactor3000\relax}%
\providecommand \BibitemShut  [1]{\csname bibitem#1\endcsname}%
\let\auto@bib@innerbib\@empty
\bibitem [{\citenamefont {{Isi}}\ \emph {et~al.}(2021)\citenamefont {{Isi}},
  \citenamefont {{Farr}}, \citenamefont {{Giesler}}, \citenamefont {{Scheel}},\
  and\ \citenamefont {{Teukolsky}}}]{2021PhRvL.127a1103I}%
  \BibitemOpen
  \bibfield  {author} {\bibinfo {author} {\bibfnamefont {M.}~\bibnamefont
  {{Isi}}}, \bibinfo {author} {\bibfnamefont {W.~M.}\ \bibnamefont {{Farr}}},
  \bibinfo {author} {\bibfnamefont {M.}~\bibnamefont {{Giesler}}}, \bibinfo
  {author} {\bibfnamefont {M.~A.}\ \bibnamefont {{Scheel}}}, \ and\ \bibinfo
  {author} {\bibfnamefont {S.~A.}\ \bibnamefont {{Teukolsky}}},\ }\href
  {\doibase 10.1103/PhysRevLett.127.011103} {\bibfield  {journal} {\bibinfo
  {journal} {\prl}\ }\textbf {\bibinfo {volume} {127}},\ \bibinfo {eid}
  {011103} (\bibinfo {year} {2021})},\ \Eprint
  {http://arxiv.org/abs/2012.04486} {arXiv:2012.04486 [gr-qc]} \BibitemShut
  {NoStop}%
\bibitem [{\citenamefont {{Cotesta}}\ \emph {et~al.}(2022)\citenamefont
  {{Cotesta}}, \citenamefont {{Carullo}}, \citenamefont {{Berti}},\ and\
  \citenamefont {{Cardoso}}}]{2022PhRvL.129k1102C}%
  \BibitemOpen
  \bibfield  {author} {\bibinfo {author} {\bibfnamefont {R.}~\bibnamefont
  {{Cotesta}}}, \bibinfo {author} {\bibfnamefont {G.}~\bibnamefont
  {{Carullo}}}, \bibinfo {author} {\bibfnamefont {E.}~\bibnamefont {{Berti}}},
  \ and\ \bibinfo {author} {\bibfnamefont {V.}~\bibnamefont {{Cardoso}}},\
  }\href {\doibase 10.1103/PhysRevLett.129.111102} {\bibfield  {journal}
  {\bibinfo  {journal} {\prl}\ }\textbf {\bibinfo {volume} {129}},\ \bibinfo
  {eid} {111102} (\bibinfo {year} {2022})},\ \Eprint
  {http://arxiv.org/abs/2201.00822} {arXiv:2201.00822 [gr-qc]} \BibitemShut
  {NoStop}%
\bibitem [{\citenamefont {{Isi}}\ and\ \citenamefont
  {{Farr}}(2022)}]{2022arXiv220202941I}%
  \BibitemOpen
  \bibfield  {author} {\bibinfo {author} {\bibfnamefont {M.}~\bibnamefont
  {{Isi}}}\ and\ \bibinfo {author} {\bibfnamefont {W.~M.}\ \bibnamefont
  {{Farr}}},\ }\href@noop {} {\bibfield  {journal} {\bibinfo  {journal} {arXiv
  e-prints}\ ,\ \bibinfo {eid} {arXiv:2202.02941}} (\bibinfo {year} {2022})},\
  \Eprint {http://arxiv.org/abs/2202.02941} {arXiv:2202.02941 [gr-qc]}
  \BibitemShut {NoStop}%
\bibitem [{\citenamefont {Abbott}\ \emph {et~al.}(2019)\citenamefont {Abbott}
  \emph {et~al.}}]{LIGO_PRX2019}%
  \BibitemOpen
  \bibfield  {author} {\bibinfo {author} {\bibfnamefont {B.~P.}\ \bibnamefont
  {Abbott}} \emph {et~al.} (\bibinfo {collaboration} {LIGO Scientific
  Collaboration and Virgo Collaboration}),\ }\href {\doibase
  10.1103/PhysRevX.9.031040} {\bibfield  {journal} {\bibinfo  {journal} {Phys.
  Rev. X}\ }\textbf {\bibinfo {volume} {9}},\ \bibinfo {pages} {031040}
  (\bibinfo {year} {2019})}\BibitemShut {NoStop}%
\bibitem [{\citenamefont {{Abbott}}\ \emph {et~al.}(2021)\citenamefont
  {{Abbott}} \emph {et~al.}}]{LIGO_O3a_PRX2020}%
  \BibitemOpen
  \bibfield  {author} {\bibinfo {author} {\bibfnamefont {R.}~\bibnamefont
  {{Abbott}}} \emph {et~al.},\ }\href {\doibase 10.1103/PhysRevX.11.021053}
  {\bibfield  {journal} {\bibinfo  {journal} {Phys. Rev. X}\ }\textbf {\bibinfo
  {volume} {11}},\ \bibinfo {eid} {021053} (\bibinfo {year} {2021})},\ \Eprint
  {http://arxiv.org/abs/2010.14527} {arXiv:2010.14527 [gr-qc]} \BibitemShut
  {NoStop}%
\bibitem [{\citenamefont {Abbott}\ \emph
  {et~al.}(2021{\natexlab{a}})\citenamefont {Abbott} \emph
  {et~al.}}]{2021arXiv211103606T}%
  \BibitemOpen
  \bibfield  {author} {\bibinfo {author} {\bibfnamefont {R.}~\bibnamefont
  {Abbott}} \emph {et~al.} (\bibinfo {collaboration} {LIGO Scientific, VIRGO,
  KAGRA}),\ }\href@noop {} {\bibfield  {journal} {\bibinfo  {journal} {arXiv
  e-prints}\ ,\ \bibinfo {eid} {arXiv:2111.03606}} (\bibinfo {year}
  {2021}{\natexlab{a}})},\ \Eprint {http://arxiv.org/abs/2111.03606}
  {arXiv:2111.03606 [gr-qc]} \BibitemShut {NoStop}%
\bibitem [{\citenamefont {Hawking}(1972)}]{Hawking:1971vc}%
  \BibitemOpen
  \bibfield  {author} {\bibinfo {author} {\bibfnamefont {S.~W.}\ \bibnamefont
  {Hawking}},\ }\href {\doibase 10.1007/BF01877517} {\bibfield  {journal}
  {\bibinfo  {journal} {Commun. Math. Phys.}\ }\textbf {\bibinfo {volume}
  {25}},\ \bibinfo {pages} {152} (\bibinfo {year} {1972})}\BibitemShut
  {NoStop}%
\bibitem [{\citenamefont {Robinson}(1975)}]{PhysRevLett.34.905}%
  \BibitemOpen
  \bibfield  {author} {\bibinfo {author} {\bibfnamefont {D.~C.}\ \bibnamefont
  {Robinson}},\ }\href {\doibase 10.1103/PhysRevLett.34.905} {\bibfield
  {journal} {\bibinfo  {journal} {Phys. Rev. Lett.}\ }\textbf {\bibinfo
  {volume} {34}},\ \bibinfo {pages} {905} (\bibinfo {year} {1975})}\BibitemShut
  {NoStop}%
\bibitem [{\citenamefont {{Vishveshwara}}(1970)}]{Schw_PRD_Vishveshwara1970}%
  \BibitemOpen
  \bibfield  {author} {\bibinfo {author} {\bibfnamefont {C.~V.}\ \bibnamefont
  {{Vishveshwara}}},\ }\href {\doibase 10.1103/PhysRevD.1.2870} {\bibfield
  {journal} {\bibinfo  {journal} {\prd}\ }\textbf {\bibinfo {volume} {1}},\
  \bibinfo {pages} {2870} (\bibinfo {year} {1970})}\BibitemShut {NoStop}%
\bibitem [{\citenamefont {{Press}}(1971)}]{GW_APJL_Press1971}%
  \BibitemOpen
  \bibfield  {author} {\bibinfo {author} {\bibfnamefont {W.~H.}\ \bibnamefont
  {{Press}}},\ }\href {\doibase 10.1086/180849} {\bibfield  {journal} {\bibinfo
   {journal} {Astrophys. J. Lett.}\ }\textbf {\bibinfo {volume} {170}},\
  \bibinfo {pages} {L105} (\bibinfo {year} {1971})}\BibitemShut {NoStop}%
\bibitem [{\citenamefont {{Teukolsky}}(1973)}]{QNM_APJ_Teukolsky1973}%
  \BibitemOpen
  \bibfield  {author} {\bibinfo {author} {\bibfnamefont {S.~A.}\ \bibnamefont
  {{Teukolsky}}},\ }\href {\doibase 10.1086/152444} {\bibfield  {journal}
  {\bibinfo  {journal} {Astrophys. J.}\ }\textbf {\bibinfo {volume} {185}},\
  \bibinfo {pages} {635} (\bibinfo {year} {1973})}\BibitemShut {NoStop}%
\bibitem [{\citenamefont {Berti}\ \emph {et~al.}(2009)\citenamefont {Berti},
  \citenamefont {Cardoso},\ and\ \citenamefont {Starinets}}]{Berti:2009kk}%
  \BibitemOpen
  \bibfield  {author} {\bibinfo {author} {\bibfnamefont {E.}~\bibnamefont
  {Berti}}, \bibinfo {author} {\bibfnamefont {V.}~\bibnamefont {Cardoso}}, \
  and\ \bibinfo {author} {\bibfnamefont {A.~O.}\ \bibnamefont {Starinets}},\
  }\href {\doibase 10.1088/0264-9381/26/16/163001} {\bibfield  {journal}
  {\bibinfo  {journal} {Class. Quant. Grav.}\ }\textbf {\bibinfo {volume}
  {26}},\ \bibinfo {pages} {163001} (\bibinfo {year} {2009})},\ \Eprint
  {http://arxiv.org/abs/0905.2975} {arXiv:0905.2975 [gr-qc]} \BibitemShut
  {NoStop}%
\bibitem [{\citenamefont {Berti}\ \emph {et~al.}(2007)\citenamefont {Berti},
  \citenamefont {Cardoso}, \citenamefont {Cardoso},\ and\ \citenamefont
  {Cavaglia}}]{Berti:2007zu}%
  \BibitemOpen
  \bibfield  {author} {\bibinfo {author} {\bibfnamefont {E.}~\bibnamefont
  {Berti}}, \bibinfo {author} {\bibfnamefont {J.}~\bibnamefont {Cardoso}},
  \bibinfo {author} {\bibfnamefont {V.}~\bibnamefont {Cardoso}}, \ and\
  \bibinfo {author} {\bibfnamefont {M.}~\bibnamefont {Cavaglia}},\ }\href
  {\doibase 10.1103/PhysRevD.76.104044} {\bibfield  {journal} {\bibinfo
  {journal} {Phys. Rev. D}\ }\textbf {\bibinfo {volume} {76}},\ \bibinfo
  {pages} {104044} (\bibinfo {year} {2007})},\ \Eprint
  {http://arxiv.org/abs/0707.1202} {arXiv:0707.1202 [gr-qc]} \BibitemShut
  {NoStop}%
\bibitem [{\citenamefont {Gossan}\ \emph {et~al.}(2012)\citenamefont {Gossan},
  \citenamefont {Veitch},\ and\ \citenamefont {Sathyaprakash}}]{Gossan:2011ha}%
  \BibitemOpen
  \bibfield  {author} {\bibinfo {author} {\bibfnamefont {S.}~\bibnamefont
  {Gossan}}, \bibinfo {author} {\bibfnamefont {J.}~\bibnamefont {Veitch}}, \
  and\ \bibinfo {author} {\bibfnamefont {B.~S.}\ \bibnamefont
  {Sathyaprakash}},\ }\href {\doibase 10.1103/PhysRevD.85.124056} {\bibfield
  {journal} {\bibinfo  {journal} {Phys. Rev. D}\ }\textbf {\bibinfo {volume}
  {85}},\ \bibinfo {pages} {124056} (\bibinfo {year} {2012})},\ \Eprint
  {http://arxiv.org/abs/1111.5819} {arXiv:1111.5819 [gr-qc]} \BibitemShut
  {NoStop}%
\bibitem [{\citenamefont {London}\ \emph {et~al.}(2014)\citenamefont {London},
  \citenamefont {Shoemaker},\ and\ \citenamefont {Healy}}]{London:2014cma}%
  \BibitemOpen
  \bibfield  {author} {\bibinfo {author} {\bibfnamefont {L.}~\bibnamefont
  {London}}, \bibinfo {author} {\bibfnamefont {D.}~\bibnamefont {Shoemaker}}, \
  and\ \bibinfo {author} {\bibfnamefont {J.}~\bibnamefont {Healy}},\ }\href
  {\doibase 10.1103/PhysRevD.90.124032} {\bibfield  {journal} {\bibinfo
  {journal} {Phys. Rev. D}\ }\textbf {\bibinfo {volume} {90}},\ \bibinfo
  {pages} {124032} (\bibinfo {year} {2014})},\ \bibinfo {note} {[Erratum:
  Phys.Rev.D 94, 069902 (2016)]},\ \Eprint {http://arxiv.org/abs/1404.3197}
  {arXiv:1404.3197 [gr-qc]} \BibitemShut {NoStop}%
\bibitem [{\citenamefont {Brito}\ \emph {et~al.}(2018)\citenamefont {Brito},
  \citenamefont {Buonanno},\ and\ \citenamefont {Raymond}}]{Brito:2018rfr}%
  \BibitemOpen
  \bibfield  {author} {\bibinfo {author} {\bibfnamefont {R.}~\bibnamefont
  {Brito}}, \bibinfo {author} {\bibfnamefont {A.}~\bibnamefont {Buonanno}}, \
  and\ \bibinfo {author} {\bibfnamefont {V.}~\bibnamefont {Raymond}},\ }\href
  {\doibase 10.1103/PhysRevD.98.084038} {\bibfield  {journal} {\bibinfo
  {journal} {Phys. Rev. D}\ }\textbf {\bibinfo {volume} {98}},\ \bibinfo
  {pages} {084038} (\bibinfo {year} {2018})},\ \Eprint
  {http://arxiv.org/abs/1805.00293} {arXiv:1805.00293 [gr-qc]} \BibitemShut
  {NoStop}%
\bibitem [{\citenamefont {Abbott}\ \emph
  {et~al.}(2016{\natexlab{a}})\citenamefont {Abbott} \emph
  {et~al.}}]{gw150914_PRL2016}%
  \BibitemOpen
  \bibfield  {author} {\bibinfo {author} {\bibfnamefont {B.~P.}\ \bibnamefont
  {Abbott}} \emph {et~al.} (\bibinfo {collaboration} {LIGO Scientific
  Collaboration and Virgo Collaboration}),\ }\href {\doibase
  10.1103/PhysRevLett.116.061102} {\bibfield  {journal} {\bibinfo  {journal}
  {Phys. Rev. Lett.}\ }\textbf {\bibinfo {volume} {116}},\ \bibinfo {pages}
  {061102} (\bibinfo {year} {2016}{\natexlab{a}})}\BibitemShut {NoStop}%
\bibitem [{\citenamefont {Abbott}\ \emph
  {et~al.}(2016{\natexlab{b}})\citenamefont {Abbott} \emph
  {et~al.}}]{LIGOScientific:2016vpg}%
  \BibitemOpen
  \bibfield  {author} {\bibinfo {author} {\bibfnamefont {B.~P.}\ \bibnamefont
  {Abbott}} \emph {et~al.} (\bibinfo {collaboration} {LIGO Scientific,
  Virgo}),\ }\href {\doibase 10.3847/2041-8205/818/2/L22} {\bibfield  {journal}
  {\bibinfo  {journal} {Astrophys. J. Lett.}\ }\textbf {\bibinfo {volume}
  {818}},\ \bibinfo {pages} {L22} (\bibinfo {year} {2016}{\natexlab{b}})},\
  \Eprint {http://arxiv.org/abs/1602.03846} {arXiv:1602.03846 [astro-ph.HE]}
  \BibitemShut {NoStop}%
\bibitem [{\citenamefont {{Isi}}\ \emph {et~al.}(2019)\citenamefont {{Isi}},
  \citenamefont {{Giesler}}, \citenamefont {{Farr}}, \citenamefont {{Scheel}},\
  and\ \citenamefont {{Teukolsky}}}]{2019PhRvL.123k1102I}%
  \BibitemOpen
  \bibfield  {author} {\bibinfo {author} {\bibfnamefont {M.}~\bibnamefont
  {{Isi}}}, \bibinfo {author} {\bibfnamefont {M.}~\bibnamefont {{Giesler}}},
  \bibinfo {author} {\bibfnamefont {W.~M.}\ \bibnamefont {{Farr}}}, \bibinfo
  {author} {\bibfnamefont {M.~A.}\ \bibnamefont {{Scheel}}}, \ and\ \bibinfo
  {author} {\bibfnamefont {S.~A.}\ \bibnamefont {{Teukolsky}}},\ }\href
  {\doibase 10.1103/PhysRevLett.123.111102} {\bibfield  {journal} {\bibinfo
  {journal} {\prl}\ }\textbf {\bibinfo {volume} {123}},\ \bibinfo {eid}
  {111102} (\bibinfo {year} {2019})},\ \Eprint
  {http://arxiv.org/abs/1905.00869} {arXiv:1905.00869 [gr-qc]} \BibitemShut
  {NoStop}%
\bibitem [{\citenamefont {Abbott}\ \emph
  {et~al.}(2021{\natexlab{b}})\citenamefont {Abbott} \emph
  {et~al.}}]{2021PhRvD.103l2002A}%
  \BibitemOpen
  \bibfield  {author} {\bibinfo {author} {\bibfnamefont {R.}~\bibnamefont
  {Abbott}} \emph {et~al.} (\bibinfo {collaboration} {LIGO Scientific,
  Virgo}),\ }\href {\doibase 10.1103/PhysRevD.103.122002} {\bibfield  {journal}
  {\bibinfo  {journal} {\prd}\ }\textbf {\bibinfo {volume} {103}},\ \bibinfo
  {eid} {122002} (\bibinfo {year} {2021}{\natexlab{b}})},\ \Eprint
  {http://arxiv.org/abs/2010.14529} {arXiv:2010.14529 [gr-qc]} \BibitemShut
  {NoStop}%
\bibitem [{\citenamefont {Abbott}\ \emph
  {et~al.}(2021{\natexlab{c}})\citenamefont {Abbott} \emph
  {et~al.}}]{2021arXiv211206861T}%
  \BibitemOpen
  \bibfield  {author} {\bibinfo {author} {\bibfnamefont {R.}~\bibnamefont
  {Abbott}} \emph {et~al.} (\bibinfo {collaboration} {LIGO Scientific, VIRGO,
  KAGRA}),\ }\href@noop {} {\bibfield  {journal} {\bibinfo  {journal} {arXiv
  e-prints}\ ,\ \bibinfo {eid} {arXiv:2112.06861}} (\bibinfo {year}
  {2021}{\natexlab{c}})},\ \Eprint {http://arxiv.org/abs/2112.06861}
  {arXiv:2112.06861 [gr-qc]} \BibitemShut {NoStop}%
\bibitem [{\citenamefont {Abbott}\ \emph
  {et~al.}(2016{\natexlab{c}})\citenamefont {Abbott} \emph
  {et~al.}}]{Test_GR_150914}%
  \BibitemOpen
  \bibfield  {author} {\bibinfo {author} {\bibfnamefont {B.~P.}\ \bibnamefont
  {Abbott}} \emph {et~al.} (\bibinfo {collaboration} {LIGO Scientific,
  Virgo}),\ }\href {\doibase 10.1103/PhysRevLett.116.221101,
  10.1103/PhysRevLett.121.129902} {\bibfield  {journal} {\bibinfo  {journal}
  {Phys. Rev. Lett.}\ }\textbf {\bibinfo {volume} {116}},\ \bibinfo {pages}
  {221101} (\bibinfo {year} {2016}{\natexlab{c}})},\ \bibinfo {note} {[Erratum:
  Phys. Rev. Lett.121,no.12,129902(2018)]},\ \Eprint
  {http://arxiv.org/abs/1602.03841} {arXiv:1602.03841 [gr-qc]} \BibitemShut
  {NoStop}%
\bibitem [{\citenamefont {Baibhav}\ \emph {et~al.}(2023)\citenamefont
  {Baibhav}, \citenamefont {Cheung}, \citenamefont {Berti}, \citenamefont
  {Cardoso}, \citenamefont {Carullo}, \citenamefont {Cotesta}, \citenamefont
  {Del~Pozzo},\ and\ \citenamefont {Duque}}]{Baibhav:2023clw}%
  \BibitemOpen
  \bibfield  {author} {\bibinfo {author} {\bibfnamefont {V.}~\bibnamefont
  {Baibhav}}, \bibinfo {author} {\bibfnamefont {M.~H.-Y.}\ \bibnamefont
  {Cheung}}, \bibinfo {author} {\bibfnamefont {E.}~\bibnamefont {Berti}},
  \bibinfo {author} {\bibfnamefont {V.}~\bibnamefont {Cardoso}}, \bibinfo
  {author} {\bibfnamefont {G.}~\bibnamefont {Carullo}}, \bibinfo {author}
  {\bibfnamefont {R.}~\bibnamefont {Cotesta}}, \bibinfo {author} {\bibfnamefont
  {W.}~\bibnamefont {Del~Pozzo}}, \ and\ \bibinfo {author} {\bibfnamefont
  {F.}~\bibnamefont {Duque}},\ }\href@noop {} {\  (\bibinfo {year} {2023})},\
  \Eprint {http://arxiv.org/abs/2302.03050} {arXiv:2302.03050 [gr-qc]}
  \BibitemShut {NoStop}%
\bibitem [{\citenamefont {Carullo}\ \emph {et~al.}(2019)\citenamefont
  {Carullo}, \citenamefont {Del~Pozzo},\ and\ \citenamefont
  {Veitch}}]{PRD_Carullo2019}%
  \BibitemOpen
  \bibfield  {author} {\bibinfo {author} {\bibfnamefont {G.}~\bibnamefont
  {Carullo}}, \bibinfo {author} {\bibfnamefont {W.}~\bibnamefont {Del~Pozzo}},
  \ and\ \bibinfo {author} {\bibfnamefont {J.}~\bibnamefont {Veitch}},\ }\href
  {\doibase 10.1103/PhysRevD.99.123029} {\bibfield  {journal} {\bibinfo
  {journal} {Phys. Rev. D}\ }\textbf {\bibinfo {volume} {99}},\ \bibinfo
  {pages} {123029} (\bibinfo {year} {2019})},\ \bibinfo {note} {[Erratum:
  Phys.Rev.D 100, 089903 (2019)]},\ \Eprint {http://arxiv.org/abs/1902.07527}
  {arXiv:1902.07527 [gr-qc]} \BibitemShut {NoStop}%
\bibitem [{\citenamefont {{Giesler}}\ \emph {et~al.}(2019)\citenamefont
  {{Giesler}}, \citenamefont {{Isi}}, \citenamefont {{Scheel}},\ and\
  \citenamefont {{Teukolsky}}}]{Overtone_PRX_Giesler2019}%
  \BibitemOpen
  \bibfield  {author} {\bibinfo {author} {\bibfnamefont {M.}~\bibnamefont
  {{Giesler}}}, \bibinfo {author} {\bibfnamefont {M.}~\bibnamefont {{Isi}}},
  \bibinfo {author} {\bibfnamefont {M.~A.}\ \bibnamefont {{Scheel}}}, \ and\
  \bibinfo {author} {\bibfnamefont {S.~A.}\ \bibnamefont {{Teukolsky}}},\
  }\href {\doibase 10.1103/PhysRevX.9.041060} {\bibfield  {journal} {\bibinfo
  {journal} {Phys. Rev. X}\ }\textbf {\bibinfo {volume} {9}},\ \bibinfo {eid}
  {041060} (\bibinfo {year} {2019})},\ \Eprint
  {http://arxiv.org/abs/1903.08284} {arXiv:1903.08284 [gr-qc]} \BibitemShut
  {NoStop}%
\bibitem [{\citenamefont {Crisostomi}\ \emph {et~al.}(2023)\citenamefont
  {Crisostomi}, \citenamefont {Dey}, \citenamefont {Barausse},\ and\
  \citenamefont {Trotta}}]{Crisostomi:2023tle}%
  \BibitemOpen
  \bibfield  {author} {\bibinfo {author} {\bibfnamefont {M.}~\bibnamefont
  {Crisostomi}}, \bibinfo {author} {\bibfnamefont {K.}~\bibnamefont {Dey}},
  \bibinfo {author} {\bibfnamefont {E.}~\bibnamefont {Barausse}}, \ and\
  \bibinfo {author} {\bibfnamefont {R.}~\bibnamefont {Trotta}},\ }\href
  {\doibase 10.1103/PhysRevD.108.044029} {\bibfield  {journal} {\bibinfo
  {journal} {Phys. Rev. D}\ }\textbf {\bibinfo {volume} {108}},\ \bibinfo
  {pages} {044029} (\bibinfo {year} {2023})},\ \Eprint
  {http://arxiv.org/abs/2305.18528} {arXiv:2305.18528 [gr-qc]} \BibitemShut
  {NoStop}%
\bibitem [{\citenamefont {Isi}\ and\ \citenamefont {Farr}(2023)}]{Isi:2023nif}%
  \BibitemOpen
  \bibfield  {author} {\bibinfo {author} {\bibfnamefont {M.}~\bibnamefont
  {Isi}}\ and\ \bibinfo {author} {\bibfnamefont {W.~M.}\ \bibnamefont {Farr}},\
  }\href {\doibase 10.1103/PhysRevLett.131.169001} {\bibfield  {journal}
  {\bibinfo  {journal} {Phys. Rev. Lett.}\ }\textbf {\bibinfo {volume} {131}},\
  \bibinfo {pages} {169001} (\bibinfo {year} {2023})},\ \Eprint
  {http://arxiv.org/abs/2310.13869} {arXiv:2310.13869 [astro-ph.HE]}
  \BibitemShut {NoStop}%
\bibitem [{\citenamefont {Carullo}\ \emph {et~al.}(2023)\citenamefont
  {Carullo}, \citenamefont {Cotesta}, \citenamefont {Berti},\ and\
  \citenamefont {Cardoso}}]{Carullo:2023gtf}%
  \BibitemOpen
  \bibfield  {author} {\bibinfo {author} {\bibfnamefont {G.}~\bibnamefont
  {Carullo}}, \bibinfo {author} {\bibfnamefont {R.}~\bibnamefont {Cotesta}},
  \bibinfo {author} {\bibfnamefont {E.}~\bibnamefont {Berti}}, \ and\ \bibinfo
  {author} {\bibfnamefont {V.}~\bibnamefont {Cardoso}},\ }\href {\doibase
  10.1103/PhysRevLett.131.169002} {\bibfield  {journal} {\bibinfo  {journal}
  {Phys. Rev. Lett.}\ }\textbf {\bibinfo {volume} {131}},\ \bibinfo {pages}
  {169002} (\bibinfo {year} {2023})},\ \Eprint
  {http://arxiv.org/abs/2310.20625} {arXiv:2310.20625 [gr-qc]} \BibitemShut
  {NoStop}%
\bibitem [{\citenamefont {{Finch}}\ and\ \citenamefont
  {{Moore}}(2021)}]{2021PhRvD.104l3034F}%
  \BibitemOpen
  \bibfield  {author} {\bibinfo {author} {\bibfnamefont {E.}~\bibnamefont
  {{Finch}}}\ and\ \bibinfo {author} {\bibfnamefont {C.~J.}\ \bibnamefont
  {{Moore}}},\ }\href {\doibase 10.1103/PhysRevD.104.123034} {\bibfield
  {journal} {\bibinfo  {journal} {\prd}\ }\textbf {\bibinfo {volume} {104}},\
  \bibinfo {eid} {123034} (\bibinfo {year} {2021})},\ \Eprint
  {http://arxiv.org/abs/2108.09344} {arXiv:2108.09344 [gr-qc]} \BibitemShut
  {NoStop}%
\bibitem [{\citenamefont {{Finch}}\ and\ \citenamefont
  {{Moore}}(2022)}]{2022PhRvD.106d3005F}%
  \BibitemOpen
  \bibfield  {author} {\bibinfo {author} {\bibfnamefont {E.}~\bibnamefont
  {{Finch}}}\ and\ \bibinfo {author} {\bibfnamefont {C.~J.}\ \bibnamefont
  {{Moore}}},\ }\href {\doibase 10.1103/PhysRevD.106.043005} {\bibfield
  {journal} {\bibinfo  {journal} {\prd}\ }\textbf {\bibinfo {volume} {106}},\
  \bibinfo {eid} {043005} (\bibinfo {year} {2022})},\ \Eprint
  {http://arxiv.org/abs/2205.07809} {arXiv:2205.07809 [gr-qc]} \BibitemShut
  {NoStop}%
\bibitem [{\citenamefont {{Bustillo}}\ \emph {et~al.}(2021)\citenamefont
  {{Bustillo}}, \citenamefont {{Lasky}},\ and\ \citenamefont
  {{Thrane}}}]{2021PhRvD.103b4041B}%
  \BibitemOpen
  \bibfield  {author} {\bibinfo {author} {\bibfnamefont {J.~C.}\ \bibnamefont
  {{Bustillo}}}, \bibinfo {author} {\bibfnamefont {P.~D.}\ \bibnamefont
  {{Lasky}}}, \ and\ \bibinfo {author} {\bibfnamefont {E.}~\bibnamefont
  {{Thrane}}},\ }\href {\doibase 10.1103/PhysRevD.103.024041} {\bibfield
  {journal} {\bibinfo  {journal} {\prd}\ }\textbf {\bibinfo {volume} {103}},\
  \bibinfo {eid} {024041} (\bibinfo {year} {2021})}\BibitemShut {NoStop}%
\bibitem [{\citenamefont {Wang}\ \emph {et~al.}(2023)\citenamefont {Wang},
  \citenamefont {Capano}, \citenamefont {Abedi}, \citenamefont {Kastha},
  \citenamefont {Krishnan}, \citenamefont {Nielsen}, \citenamefont {Nitz},\
  and\ \citenamefont {Westerweck}}]{Wang:2023xsy}%
  \BibitemOpen
  \bibfield  {author} {\bibinfo {author} {\bibfnamefont {Y.-F.}\ \bibnamefont
  {Wang}}, \bibinfo {author} {\bibfnamefont {C.~D.}\ \bibnamefont {Capano}},
  \bibinfo {author} {\bibfnamefont {J.}~\bibnamefont {Abedi}}, \bibinfo
  {author} {\bibfnamefont {S.}~\bibnamefont {Kastha}}, \bibinfo {author}
  {\bibfnamefont {B.}~\bibnamefont {Krishnan}}, \bibinfo {author}
  {\bibfnamefont {A.~B.}\ \bibnamefont {Nielsen}}, \bibinfo {author}
  {\bibfnamefont {A.~H.}\ \bibnamefont {Nitz}}, \ and\ \bibinfo {author}
  {\bibfnamefont {J.}~\bibnamefont {Westerweck}},\ }\href@noop {} {\  (\bibinfo
  {year} {2023})},\ \Eprint {http://arxiv.org/abs/2310.19645} {arXiv:2310.19645
  [gr-qc]} \BibitemShut {NoStop}%
\bibitem [{\citenamefont {{Ma}}\ \emph {et~al.}(2022)\citenamefont {{Ma}},
  \citenamefont {{Mitman}}, \citenamefont {{Sun}}, \citenamefont {{Deppe}},
  \citenamefont {{H{\'e}bert}}, \citenamefont {{Kidder}}, \citenamefont
  {{Moxon}}, \citenamefont {{Throwe}}, \citenamefont {{Vu}},\ and\
  \citenamefont {{Chen}}}]{2022PhRvD.106h4036M}%
  \BibitemOpen
  \bibfield  {author} {\bibinfo {author} {\bibfnamefont {S.}~\bibnamefont
  {{Ma}}}, \bibinfo {author} {\bibfnamefont {K.}~\bibnamefont {{Mitman}}},
  \bibinfo {author} {\bibfnamefont {L.}~\bibnamefont {{Sun}}}, \bibinfo
  {author} {\bibfnamefont {N.}~\bibnamefont {{Deppe}}}, \bibinfo {author}
  {\bibfnamefont {F.}~\bibnamefont {{H{\'e}bert}}}, \bibinfo {author}
  {\bibfnamefont {L.~E.}\ \bibnamefont {{Kidder}}}, \bibinfo {author}
  {\bibfnamefont {J.}~\bibnamefont {{Moxon}}}, \bibinfo {author} {\bibfnamefont
  {W.}~\bibnamefont {{Throwe}}}, \bibinfo {author} {\bibfnamefont {N.~L.}\
  \bibnamefont {{Vu}}}, \ and\ \bibinfo {author} {\bibfnamefont
  {Y.}~\bibnamefont {{Chen}}},\ }\href {\doibase 10.1103/PhysRevD.106.084036}
  {\bibfield  {journal} {\bibinfo  {journal} {\prd}\ }\textbf {\bibinfo
  {volume} {106}},\ \bibinfo {eid} {084036} (\bibinfo {year} {2022})},\ \Eprint
  {http://arxiv.org/abs/2207.10870} {arXiv:2207.10870 [gr-qc]} \BibitemShut
  {NoStop}%
\bibitem [{\citenamefont {Ma}\ \emph {et~al.}(2023{\natexlab{a}})\citenamefont
  {Ma}, \citenamefont {Sun},\ and\ \citenamefont {Chen}}]{Ma:2023vvr}%
  \BibitemOpen
  \bibfield  {author} {\bibinfo {author} {\bibfnamefont {S.}~\bibnamefont
  {Ma}}, \bibinfo {author} {\bibfnamefont {L.}~\bibnamefont {Sun}}, \ and\
  \bibinfo {author} {\bibfnamefont {Y.}~\bibnamefont {Chen}},\ }\href {\doibase
  10.1103/PhysRevD.107.084010} {\bibfield  {journal} {\bibinfo  {journal}
  {Phys. Rev. D}\ }\textbf {\bibinfo {volume} {107}},\ \bibinfo {pages}
  {084010} (\bibinfo {year} {2023}{\natexlab{a}})},\ \Eprint
  {http://arxiv.org/abs/2301.06639} {arXiv:2301.06639 [gr-qc]} \BibitemShut
  {NoStop}%
\bibitem [{\citenamefont {Ma}\ \emph {et~al.}(2023{\natexlab{b}})\citenamefont
  {Ma}, \citenamefont {Sun},\ and\ \citenamefont {Chen}}]{Ma:2023cwe}%
  \BibitemOpen
  \bibfield  {author} {\bibinfo {author} {\bibfnamefont {S.}~\bibnamefont
  {Ma}}, \bibinfo {author} {\bibfnamefont {L.}~\bibnamefont {Sun}}, \ and\
  \bibinfo {author} {\bibfnamefont {Y.}~\bibnamefont {Chen}},\ }\href {\doibase
  10.1103/PhysRevLett.130.141401} {\bibfield  {journal} {\bibinfo  {journal}
  {Phys. Rev. Lett.}\ }\textbf {\bibinfo {volume} {130}},\ \bibinfo {pages}
  {141401} (\bibinfo {year} {2023}{\natexlab{b}})},\ \Eprint
  {http://arxiv.org/abs/2301.06705} {arXiv:2301.06705 [gr-qc]} \BibitemShut
  {NoStop}%
\bibitem [{\citenamefont {{Wang}}\ \emph {et~al.}(2021)\citenamefont {{Wang}},
  \citenamefont {{Tang}}, \citenamefont {{Li}},\ and\ \citenamefont
  {{Fan}}}]{2021PhRvD.104j4063W}%
  \BibitemOpen
  \bibfield  {author} {\bibinfo {author} {\bibfnamefont {H.-T.}\ \bibnamefont
  {{Wang}}}, \bibinfo {author} {\bibfnamefont {S.-P.}\ \bibnamefont {{Tang}}},
  \bibinfo {author} {\bibfnamefont {P.-C.}\ \bibnamefont {{Li}}}, \ and\
  \bibinfo {author} {\bibfnamefont {Y.-Z.}\ \bibnamefont {{Fan}}},\ }\href
  {\doibase 10.1103/PhysRevD.104.104063} {\bibfield  {journal} {\bibinfo
  {journal} {\prd}\ }\textbf {\bibinfo {volume} {104}},\ \bibinfo {eid}
  {104063} (\bibinfo {year} {2021})},\ \Eprint
  {http://arxiv.org/abs/2104.07594} {arXiv:2104.07594 [gr-qc]} \BibitemShut
  {NoStop}%
\bibitem [{\citenamefont {Cheung}\ \emph {et~al.}(2021)\citenamefont {Cheung},
  \citenamefont {Poon}, \citenamefont {Chung},\ and\ \citenamefont
  {Li}}]{Cheung:2020dxo}%
  \BibitemOpen
  \bibfield  {author} {\bibinfo {author} {\bibfnamefont {M.~H.-Y.}\
  \bibnamefont {Cheung}}, \bibinfo {author} {\bibfnamefont {L.~W.-H.}\
  \bibnamefont {Poon}}, \bibinfo {author} {\bibfnamefont {A.~K.-W.}\
  \bibnamefont {Chung}}, \ and\ \bibinfo {author} {\bibfnamefont {T.~G.~F.}\
  \bibnamefont {Li}},\ }\href {\doibase 10.1088/1475-7516/2021/02/040}
  {\bibfield  {journal} {\bibinfo  {journal} {JCAP}\ }\textbf {\bibinfo
  {volume} {02}},\ \bibinfo {pages} {040} (\bibinfo {year} {2021})},\ \Eprint
  {http://arxiv.org/abs/2002.01695} {arXiv:2002.01695 [gr-qc]} \BibitemShut
  {NoStop}%
\bibitem [{\citenamefont {Mishra}\ \emph {et~al.}(2022)\citenamefont {Mishra},
  \citenamefont {Ghosh},\ and\ \citenamefont {Chakraborty}}]{Mishra:2021waw}%
  \BibitemOpen
  \bibfield  {author} {\bibinfo {author} {\bibfnamefont {A.~K.}\ \bibnamefont
  {Mishra}}, \bibinfo {author} {\bibfnamefont {A.}~\bibnamefont {Ghosh}}, \
  and\ \bibinfo {author} {\bibfnamefont {S.}~\bibnamefont {Chakraborty}},\
  }\href {\doibase 10.1140/epjc/s10052-022-10788-x} {\bibfield  {journal}
  {\bibinfo  {journal} {Eur. Phys. J. C}\ }\textbf {\bibinfo {volume} {82}},\
  \bibinfo {pages} {820} (\bibinfo {year} {2022})},\ \Eprint
  {http://arxiv.org/abs/2106.05558} {arXiv:2106.05558 [gr-qc]} \BibitemShut
  {NoStop}%
\bibitem [{\citenamefont {Carullo}(2021)}]{Carullo:2021dui}%
  \BibitemOpen
  \bibfield  {author} {\bibinfo {author} {\bibfnamefont {G.}~\bibnamefont
  {Carullo}},\ }\href {\doibase 10.1103/PhysRevD.103.124043} {\bibfield
  {journal} {\bibinfo  {journal} {Phys. Rev. D}\ }\textbf {\bibinfo {volume}
  {103}},\ \bibinfo {pages} {124043} (\bibinfo {year} {2021})},\ \Eprint
  {http://arxiv.org/abs/2102.05939} {arXiv:2102.05939 [gr-qc]} \BibitemShut
  {NoStop}%
\bibitem [{\citenamefont {Dey}\ \emph {et~al.}(2023)\citenamefont {Dey},
  \citenamefont {Barausse},\ and\ \citenamefont {Basak}}]{Dey:2022pmv}%
  \BibitemOpen
  \bibfield  {author} {\bibinfo {author} {\bibfnamefont {K.}~\bibnamefont
  {Dey}}, \bibinfo {author} {\bibfnamefont {E.}~\bibnamefont {Barausse}}, \
  and\ \bibinfo {author} {\bibfnamefont {S.}~\bibnamefont {Basak}},\ }\href
  {\doibase 10.1103/PhysRevD.108.024064} {\bibfield  {journal} {\bibinfo
  {journal} {Phys. Rev. D}\ }\textbf {\bibinfo {volume} {108}},\ \bibinfo
  {pages} {024064} (\bibinfo {year} {2023})},\ \Eprint
  {http://arxiv.org/abs/2212.10725} {arXiv:2212.10725 [gr-qc]} \BibitemShut
  {NoStop}%
\bibitem [{\citenamefont {Carullo}\ \emph {et~al.}(2022)\citenamefont
  {Carullo}, \citenamefont {Laghi}, \citenamefont {Johnson-McDaniel},
  \citenamefont {Del~Pozzo}, \citenamefont {Dias}, \citenamefont {Godazgar},\
  and\ \citenamefont {Santos}}]{Carullo:2021oxn}%
  \BibitemOpen
  \bibfield  {author} {\bibinfo {author} {\bibfnamefont {G.}~\bibnamefont
  {Carullo}}, \bibinfo {author} {\bibfnamefont {D.}~\bibnamefont {Laghi}},
  \bibinfo {author} {\bibfnamefont {N.~K.}\ \bibnamefont {Johnson-McDaniel}},
  \bibinfo {author} {\bibfnamefont {W.}~\bibnamefont {Del~Pozzo}}, \bibinfo
  {author} {\bibfnamefont {O.~J.~C.}\ \bibnamefont {Dias}}, \bibinfo {author}
  {\bibfnamefont {M.}~\bibnamefont {Godazgar}}, \ and\ \bibinfo {author}
  {\bibfnamefont {J.~E.}\ \bibnamefont {Santos}},\ }\href {\doibase
  10.1103/PhysRevD.105.062009} {\bibfield  {journal} {\bibinfo  {journal}
  {Phys. Rev. D}\ }\textbf {\bibinfo {volume} {105}},\ \bibinfo {pages}
  {062009} (\bibinfo {year} {2022})},\ \Eprint
  {http://arxiv.org/abs/2109.13961} {arXiv:2109.13961 [gr-qc]} \BibitemShut
  {NoStop}%
\bibitem [{\citenamefont {Gu}\ \emph {et~al.}(2023)\citenamefont {Gu},
  \citenamefont {Wang},\ and\ \citenamefont {Shao}}]{Gu:2023eaa}%
  \BibitemOpen
  \bibfield  {author} {\bibinfo {author} {\bibfnamefont {H.-P.}\ \bibnamefont
  {Gu}}, \bibinfo {author} {\bibfnamefont {H.-T.}\ \bibnamefont {Wang}}, \ and\
  \bibinfo {author} {\bibfnamefont {L.}~\bibnamefont {Shao}},\ }\href@noop {}
  {\  (\bibinfo {year} {2023})},\ \Eprint {http://arxiv.org/abs/2310.10447}
  {arXiv:2310.10447 [gr-qc]} \BibitemShut {NoStop}%
\bibitem [{\citenamefont {Laghi}\ \emph {et~al.}(2021)\citenamefont {Laghi},
  \citenamefont {Carullo}, \citenamefont {Veitch},\ and\ \citenamefont
  {Del~Pozzo}}]{Laghi:2020rgl}%
  \BibitemOpen
  \bibfield  {author} {\bibinfo {author} {\bibfnamefont {D.}~\bibnamefont
  {Laghi}}, \bibinfo {author} {\bibfnamefont {G.}~\bibnamefont {Carullo}},
  \bibinfo {author} {\bibfnamefont {J.}~\bibnamefont {Veitch}}, \ and\ \bibinfo
  {author} {\bibfnamefont {W.}~\bibnamefont {Del~Pozzo}},\ }\href {\doibase
  10.1088/1361-6382/abde19} {\bibfield  {journal} {\bibinfo  {journal} {Class.
  Quant. Grav.}\ }\textbf {\bibinfo {volume} {38}},\ \bibinfo {pages} {095005}
  (\bibinfo {year} {2021})},\ \Eprint {http://arxiv.org/abs/2011.03816}
  {arXiv:2011.03816 [gr-qc]} \BibitemShut {NoStop}%
\bibitem [{\citenamefont {Hu}\ \emph {et~al.}(2021)\citenamefont {Hu},
  \citenamefont {Jani}, \citenamefont {Holley-Bockelmann},\ and\ \citenamefont
  {Carullo}}]{Hu:2021lbt}%
  \BibitemOpen
  \bibfield  {author} {\bibinfo {author} {\bibfnamefont {P.}~\bibnamefont
  {Hu}}, \bibinfo {author} {\bibfnamefont {K.}~\bibnamefont {Jani}}, \bibinfo
  {author} {\bibfnamefont {K.}~\bibnamefont {Holley-Bockelmann}}, \ and\
  \bibinfo {author} {\bibfnamefont {G.}~\bibnamefont {Carullo}},\ }\href@noop
  {} {\  (\bibinfo {year} {2021})},\ \Eprint {http://arxiv.org/abs/2112.06856}
  {arXiv:2112.06856 [gr-qc]} \BibitemShut {NoStop}%
\bibitem [{\citenamefont {Carullo}\ \emph {et~al.}(2021)\citenamefont
  {Carullo}, \citenamefont {Laghi}, \citenamefont {Veitch},\ and\ \citenamefont
  {Del~Pozzo}}]{Carullo:2021yxh}%
  \BibitemOpen
  \bibfield  {author} {\bibinfo {author} {\bibfnamefont {G.}~\bibnamefont
  {Carullo}}, \bibinfo {author} {\bibfnamefont {D.}~\bibnamefont {Laghi}},
  \bibinfo {author} {\bibfnamefont {J.}~\bibnamefont {Veitch}}, \ and\ \bibinfo
  {author} {\bibfnamefont {W.}~\bibnamefont {Del~Pozzo}},\ }\href {\doibase
  10.1103/PhysRevLett.126.161102} {\bibfield  {journal} {\bibinfo  {journal}
  {Phys. Rev. Lett.}\ }\textbf {\bibinfo {volume} {126}},\ \bibinfo {pages}
  {161102} (\bibinfo {year} {2021})},\ \Eprint
  {http://arxiv.org/abs/2103.06167} {arXiv:2103.06167 [gr-qc]} \BibitemShut
  {NoStop}%
\bibitem [{\citenamefont {{Isi}}\ and\ \citenamefont
  {{Farr}}(2021)}]{2021arXiv210705609I}%
  \BibitemOpen
  \bibfield  {author} {\bibinfo {author} {\bibfnamefont {M.}~\bibnamefont
  {{Isi}}}\ and\ \bibinfo {author} {\bibfnamefont {W.~M.}\ \bibnamefont
  {{Farr}}},\ }\href@noop {} {\bibfield  {journal} {\bibinfo  {journal} {arXiv
  e-prints}\ ,\ \bibinfo {eid} {arXiv:2107.05609}} (\bibinfo {year} {2021})},\
  \Eprint {http://arxiv.org/abs/2107.05609} {arXiv:2107.05609 [gr-qc]}
  \BibitemShut {NoStop}%
\bibitem [{\citenamefont {Khan}\ and\ \citenamefont
  {Agha}(2020)}]{KhanFIR2020}%
  \BibitemOpen
  \bibfield  {author} {\bibinfo {author} {\bibfnamefont {M.}~\bibnamefont
  {Khan}}\ and\ \bibinfo {author} {\bibfnamefont {S.}~\bibnamefont {Agha}},\
  }\href {\doibase 10.1007/s10470-020-01688-9} {\bibfield  {journal} {\bibinfo
  {journal} {Analog Integrated Circuits and Signal Processing}\ }\textbf
  {\bibinfo {volume} {105}},\ \bibinfo {pages} {99} (\bibinfo {year}
  {2020})}\BibitemShut {NoStop}%
\bibitem [{\citenamefont {{Allen}}\ \emph {et~al.}(2012)\citenamefont
  {{Allen}}, \citenamefont {{Anderson}}, \citenamefont {{Brady}}, \citenamefont
  {{Brown}},\ and\ \citenamefont {{Creighton}}}]{2012PhRvD..85l2006A}%
  \BibitemOpen
  \bibfield  {author} {\bibinfo {author} {\bibfnamefont {B.}~\bibnamefont
  {{Allen}}}, \bibinfo {author} {\bibfnamefont {W.~G.}\ \bibnamefont
  {{Anderson}}}, \bibinfo {author} {\bibfnamefont {P.~R.}\ \bibnamefont
  {{Brady}}}, \bibinfo {author} {\bibfnamefont {D.~A.}\ \bibnamefont
  {{Brown}}}, \ and\ \bibinfo {author} {\bibfnamefont {J.~D.~E.}\ \bibnamefont
  {{Creighton}}},\ }\href {\doibase 10.1103/PhysRevD.85.122006} {\bibfield
  {journal} {\bibinfo  {journal} {\prd}\ }\textbf {\bibinfo {volume} {85}},\
  \bibinfo {eid} {122006} (\bibinfo {year} {2012})},\ \Eprint
  {http://arxiv.org/abs/gr-qc/0509116} {arXiv:gr-qc/0509116 [gr-qc]}
  \BibitemShut {NoStop}%
\bibitem [{\citenamefont {Scientific}\ and\ \citenamefont
  {Collaborations}(2019)}]{psd:gwtc1}%
  \BibitemOpen
  \bibfield  {author} {\bibinfo {author} {\bibfnamefont {L.}~\bibnamefont
  {Scientific}}\ and\ \bibinfo {author} {\bibfnamefont {V.}~\bibnamefont
  {Collaborations}} (\bibinfo {collaboration} {LIGO Scientific, Virgo}),\
  }\href {\doibase 10.7935/KSX7-QQ51} {\  (\bibinfo {year} {2019}),\
  10.7935/KSX7-QQ51}\BibitemShut {NoStop}%
\bibitem [{\citenamefont {Welch}(1967)}]{1967D.Welch}%
  \BibitemOpen
  \bibfield  {author} {\bibinfo {author} {\bibfnamefont {P.~D.}\ \bibnamefont
  {Welch}},\ }\href {\doibase 10.1109/TAU.1967.1161901} {\bibfield  {journal}
  {\bibinfo  {journal} {IEEE Trans. Audio \& Electroacoust}\ }\textbf {\bibinfo
  {volume} {15}} (\bibinfo {year} {1967}),\
  10.1109/TAU.1967.1161901}\BibitemShut {NoStop}%
\bibitem [{\citenamefont {Littenberg}\ and\ \citenamefont
  {Cornish}(2015)}]{Littenberg:2014oda}%
  \BibitemOpen
  \bibfield  {author} {\bibinfo {author} {\bibfnamefont {T.~B.}\ \bibnamefont
  {Littenberg}}\ and\ \bibinfo {author} {\bibfnamefont {N.~J.}\ \bibnamefont
  {Cornish}},\ }\href {\doibase 10.1103/PhysRevD.91.084034} {\bibfield
  {journal} {\bibinfo  {journal} {Phys. Rev. D}\ }\textbf {\bibinfo {volume}
  {91}},\ \bibinfo {pages} {084034} (\bibinfo {year} {2015})},\ \Eprint
  {http://arxiv.org/abs/1410.3852} {arXiv:1410.3852 [gr-qc]} \BibitemShut
  {NoStop}%
\bibitem [{\citenamefont {Cornish}\ and\ \citenamefont
  {Littenberg}(2015)}]{Cornish:2014kda}%
  \BibitemOpen
  \bibfield  {author} {\bibinfo {author} {\bibfnamefont {N.~J.}\ \bibnamefont
  {Cornish}}\ and\ \bibinfo {author} {\bibfnamefont {T.~B.}\ \bibnamefont
  {Littenberg}},\ }\href {\doibase 10.1088/0264-9381/32/13/135012} {\bibfield
  {journal} {\bibinfo  {journal} {Class. Quant. Grav.}\ }\textbf {\bibinfo
  {volume} {32}},\ \bibinfo {pages} {135012} (\bibinfo {year} {2015})},\
  \Eprint {http://arxiv.org/abs/1410.3835} {arXiv:1410.3835 [gr-qc]}
  \BibitemShut {NoStop}%
\bibitem [{\citenamefont {Scientific}\ and\ \citenamefont
  {Collaborations}(2020)}]{data:gw150914}%
  \BibitemOpen
  \bibfield  {author} {\bibinfo {author} {\bibfnamefont {L.}~\bibnamefont
  {Scientific}}\ and\ \bibinfo {author} {\bibfnamefont {V.}~\bibnamefont
  {Collaborations}} (\bibinfo {collaboration} {LIGO Scientific, Virgo}),\
  }\href {\doibase 10.7935/82H3-HH23} {\  (\bibinfo {year} {2020}),\
  10.7935/82H3-HH23}\BibitemShut {NoStop}%
\bibitem [{\citenamefont {Moore}\ \emph {et~al.}(2015)\citenamefont {Moore},
  \citenamefont {Cole},\ and\ \citenamefont {Berry}}]{Moore:2014lga}%
  \BibitemOpen
  \bibfield  {author} {\bibinfo {author} {\bibfnamefont {C.~J.}\ \bibnamefont
  {Moore}}, \bibinfo {author} {\bibfnamefont {R.~H.}\ \bibnamefont {Cole}}, \
  and\ \bibinfo {author} {\bibfnamefont {C.~P.~L.}\ \bibnamefont {Berry}},\
  }\href {\doibase 10.1088/0264-9381/32/1/015014} {\bibfield  {journal}
  {\bibinfo  {journal} {Class. Quant. Grav.}\ }\textbf {\bibinfo {volume}
  {32}},\ \bibinfo {pages} {015014} (\bibinfo {year} {2015})},\ \Eprint
  {http://arxiv.org/abs/1408.0740} {arXiv:1408.0740 [gr-qc]} \BibitemShut
  {NoStop}%
\bibitem [{\citenamefont {Ashton}\ \emph {et~al.}(2019)\citenamefont {Ashton}
  \emph {et~al.}}]{Ashton_APJ2019}%
  \BibitemOpen
  \bibfield  {author} {\bibinfo {author} {\bibfnamefont {G.}~\bibnamefont
  {Ashton}} \emph {et~al.},\ }\href {\doibase 10.3847/1538-4365/ab06fc}
  {\bibfield  {journal} {\bibinfo  {journal} {ApJS}\ }\textbf {\bibinfo
  {volume} {241}},\ \bibinfo {pages} {27} (\bibinfo {year} {2019})}\BibitemShut
  {NoStop}%
\bibitem [{\citenamefont {{Speagle}}(2020)}]{Dynesty_MNRAS_Speagle2020}%
  \BibitemOpen
  \bibfield  {author} {\bibinfo {author} {\bibfnamefont {J.~S.}\ \bibnamefont
  {{Speagle}}},\ }\href {\doibase 10.1093/mnras/staa278} {\bibfield  {journal}
  {\bibinfo  {journal} {\mnras}\ }\textbf {\bibinfo {volume} {493}},\ \bibinfo
  {pages} {3132} (\bibinfo {year} {2020})},\ \Eprint
  {http://arxiv.org/abs/1904.02180} {arXiv:1904.02180 [astro-ph.IM]}
  \BibitemShut {NoStop}%
\bibitem [{\citenamefont {Nee}\ \emph {et~al.}(2023)\citenamefont {Nee},
  \citenamefont {V\"olkel},\ and\ \citenamefont {Pfeiffer}}]{Nee:2023osy}%
  \BibitemOpen
  \bibfield  {author} {\bibinfo {author} {\bibfnamefont {P.~J.}\ \bibnamefont
  {Nee}}, \bibinfo {author} {\bibfnamefont {S.~H.}\ \bibnamefont {V\"olkel}}, \
  and\ \bibinfo {author} {\bibfnamefont {H.~P.}\ \bibnamefont {Pfeiffer}},\
  }\href {\doibase 10.1103/PhysRevD.108.044032} {\bibfield  {journal} {\bibinfo
   {journal} {Phys. Rev. D}\ }\textbf {\bibinfo {volume} {108}},\ \bibinfo
  {pages} {044032} (\bibinfo {year} {2023})},\ \Eprint
  {http://arxiv.org/abs/2302.06634} {arXiv:2302.06634 [gr-qc]} \BibitemShut
  {NoStop}%
\bibitem [{\citenamefont {Khera}\ \emph {et~al.}(2023)\citenamefont {Khera},
  \citenamefont {Ribes~Metidieri}, \citenamefont {Bonga}, \citenamefont
  {Forteza}, \citenamefont {Krishnan}, \citenamefont {Poisson}, \citenamefont
  {Pook-Kolb}, \citenamefont {Schnetter},\ and\ \citenamefont
  {Yang}}]{Khera:2023lnc}%
  \BibitemOpen
  \bibfield  {author} {\bibinfo {author} {\bibfnamefont {N.}~\bibnamefont
  {Khera}}, \bibinfo {author} {\bibfnamefont {A.}~\bibnamefont
  {Ribes~Metidieri}}, \bibinfo {author} {\bibfnamefont {B.}~\bibnamefont
  {Bonga}}, \bibinfo {author} {\bibfnamefont {X.~J.}\ \bibnamefont {Forteza}},
  \bibinfo {author} {\bibfnamefont {B.}~\bibnamefont {Krishnan}}, \bibinfo
  {author} {\bibfnamefont {E.}~\bibnamefont {Poisson}}, \bibinfo {author}
  {\bibfnamefont {D.}~\bibnamefont {Pook-Kolb}}, \bibinfo {author}
  {\bibfnamefont {E.}~\bibnamefont {Schnetter}}, \ and\ \bibinfo {author}
  {\bibfnamefont {H.}~\bibnamefont {Yang}},\ }\href@noop {} {\  (\bibinfo
  {year} {2023})},\ \Eprint {http://arxiv.org/abs/2306.11142} {arXiv:2306.11142
  [gr-qc]} \BibitemShut {NoStop}%
\bibitem [{\citenamefont {{Punturo}}\ \emph {et~al.}(2010)\citenamefont
  {{Punturo}}, \citenamefont {{Abernathy}} \emph
  {et~al.}}]{2010CQGra..27s4002P}%
  \BibitemOpen
  \bibfield  {author} {\bibinfo {author} {\bibfnamefont {M.}~\bibnamefont
  {{Punturo}}}, \bibinfo {author} {\bibfnamefont {M.}~\bibnamefont
  {{Abernathy}}},  \emph {et~al.},\ }\href {\doibase
  10.1088/0264-9381/27/19/194002} {\bibfield  {journal} {\bibinfo  {journal}
  {Class. Quantum Grav.}\ }\textbf {\bibinfo {volume} {27}},\ \bibinfo {eid}
  {194002} (\bibinfo {year} {2010})}\BibitemShut {NoStop}%
\bibitem [{\citenamefont {{Reitze}}\ \emph {et~al.}(2019)\citenamefont
  {{Reitze}}, \citenamefont {{Adhikari}} \emph {et~al.}}]{2019BAAS...51g..35R}%
  \BibitemOpen
  \bibfield  {author} {\bibinfo {author} {\bibfnamefont {D.}~\bibnamefont
  {{Reitze}}}, \bibinfo {author} {\bibfnamefont {R.~X.}\ \bibnamefont
  {{Adhikari}}},  \emph {et~al.},\ }in\ \href@noop {} {\emph {\bibinfo
  {booktitle} {Bulletin of the American Astronomical Society}}},\ Vol.~\bibinfo
  {volume} {51}\ (\bibinfo {year} {2019})\ p.~\bibinfo {pages} {35},\ \Eprint
  {http://arxiv.org/abs/1907.04833} {arXiv:1907.04833 [astro-ph.IM]}
  \BibitemShut {NoStop}%
\bibitem [{\citenamefont {{Amaro-Seoane}}\ \emph {et~al.}(2017)\citenamefont
  {{Amaro-Seoane}}, \citenamefont {{Audley}}, \citenamefont {{Babak}},
  \citenamefont {{Baker}} \emph {et~al.}}]{LISA_arxiv2017}%
  \BibitemOpen
  \bibfield  {author} {\bibinfo {author} {\bibfnamefont {P.}~\bibnamefont
  {{Amaro-Seoane}}}, \bibinfo {author} {\bibfnamefont {H.}~\bibnamefont
  {{Audley}}}, \bibinfo {author} {\bibfnamefont {S.}~\bibnamefont {{Babak}}},
  \bibinfo {author} {\bibfnamefont {J.}~\bibnamefont {{Baker}}},  \emph
  {et~al.},\ }\href@noop {} {\bibfield  {journal} {\bibinfo  {journal} {ArXiv
  e-prints}\ ,\ \bibinfo {eid} {arXiv:1702.00786}} (\bibinfo {year} {2017})},\
  \Eprint {http://arxiv.org/abs/1702.00786} {arXiv:1702.00786 [astro-ph.IM]}
  \BibitemShut {NoStop}%
\bibitem [{\citenamefont {Luo}\ \emph {et~al.}(2016)\citenamefont {Luo} \emph
  {et~al.}}]{TQ_2015}%
  \BibitemOpen
  \bibfield  {author} {\bibinfo {author} {\bibfnamefont {J.}~\bibnamefont
  {Luo}} \emph {et~al.} (\bibinfo {collaboration} {TianQin}),\ }\href {\doibase
  10.1088/0264-9381/33/3/035010} {\bibfield  {journal} {\bibinfo  {journal}
  {Class. Quant. Grav.}\ }\textbf {\bibinfo {volume} {33}},\ \bibinfo {pages}
  {035010} (\bibinfo {year} {2016})},\ \Eprint
  {http://arxiv.org/abs/1512.02076} {arXiv:1512.02076 [astro-ph.IM]}
  \BibitemShut {NoStop}%
\bibitem [{\citenamefont {Mei}\ \emph {et~al.}(2021)\citenamefont {Mei} \emph
  {et~al.}}]{TianQin:2020hid}%
  \BibitemOpen
  \bibfield  {author} {\bibinfo {author} {\bibfnamefont {J.}~\bibnamefont
  {Mei}} \emph {et~al.} (\bibinfo {collaboration} {TianQin}),\ }\href {\doibase
  10.1093/ptep/ptaa114} {\bibfield  {journal} {\bibinfo  {journal} {PTEP}\
  }\textbf {\bibinfo {volume} {2021}},\ \bibinfo {pages} {05A107} (\bibinfo
  {year} {2021})},\ \Eprint {http://arxiv.org/abs/2008.10332} {arXiv:2008.10332
  [gr-qc]} \BibitemShut {NoStop}%
\bibitem [{\citenamefont {Hu}\ and\ \citenamefont {Wu}(2017)}]{1093nsrnwx116}%
  \BibitemOpen
  \bibfield  {author} {\bibinfo {author} {\bibfnamefont {W.-R.}\ \bibnamefont
  {Hu}}\ and\ \bibinfo {author} {\bibfnamefont {Y.-L.}\ \bibnamefont {Wu}},\
  }\href {\doibase 10.1093/nsr/nwx116} {\bibfield  {journal} {\bibinfo
  {journal} {National Science Review}\ }\textbf {\bibinfo {volume} {4}},\
  \bibinfo {pages} {685} (\bibinfo {year} {2017})}\BibitemShut {NoStop}%
\bibitem [{\citenamefont {Wang}\ and\ \citenamefont
  {Shao}(2023)}]{acf_estimation_ttd2}%
  \BibitemOpen
  \bibfield  {author} {\bibinfo {author} {\bibfnamefont {H.}~\bibnamefont
  {Wang}}\ and\ \bibinfo {author} {\bibfnamefont {L.}~\bibnamefont {Shao}},\
  }\href {https://github.com/whaitian/TTD2.git} {\bibfield  {journal} {\bibinfo
   {journal} {GitHub}\ } (\bibinfo {year} {2023})}\BibitemShut {NoStop}%
\bibitem [{\citenamefont {Abbott}\ \emph {et~al.}(2023)\citenamefont {Abbott}
  \emph {et~al.}}]{LIGOScientific:2023vdi}%
  \BibitemOpen
  \bibfield  {author} {\bibinfo {author} {\bibfnamefont {R.}~\bibnamefont
  {Abbott}} \emph {et~al.} (\bibinfo {collaboration} {LIGO Scientific, VIRGO,
  KAGRA}),\ }\href@noop {} {\  (\bibinfo {year} {2023})},\ \Eprint
  {http://arxiv.org/abs/2302.03676} {arXiv:2302.03676 [gr-qc]} \BibitemShut
  {NoStop}%
\end{thebibliography}%

\end{document}